# A Simple Condensation Model for the $H_2SO_4$-$H_2O$ Gas-cloud System on Venus

Longkang Dai[1], Xi Zhang[2], Wencheng D. Shao[2], Carver J. Bierson[3], Jun Cui[1,4]

[1]Planetary Environmental and Astrobiological Research Laboratory (PEARL), School of Atmospheric Sciences, Sun Yat-sen University, Zhuhai, Guangdong 519082, China;
[2]Department of Earth and Planetary Sciences, UC Santa Cruz, Santa Cruz, CA 95064, USA;
[3]School of Earth and Space Exploration, Arizona State University, Tempe, AZ 85281, USA
[4]Chinese Academy of Sciences Center for Excellence in Comparative Planetology, Hefei, Anhui 230026, China

Corresponding author: Longkang Dai (dailk@mail2.sysu.edu.cn)

**Key Points:**

- A simple cloud condensation model without detailed microphysics can explain the observed vertical structures of vapor and clouds on Venus.

- The simulated sulfuric acid vapor is largely supersaturated above 60 km at altitudes of $H_2SO_4$ chemical production.

- The sulfuric acid vapor flux recycled in the lower clouds is seven times larger than the downward flux of sulfuric acid in the upper clouds.

**Abstract**

The current Venus climate is largely regulated by globally-covered concentrated sulfuric acid clouds from binary condensation of sulfuric acid ($H_2SO_4$) and water ($H_2O$). To understand this complicated $H_2SO_4$-$H_2O$ gas-cloud system, previous theoretical studies either adopted complicated microphysical calculations or assumed that both $H_2SO_4$ and $H_2O$ vapor follow their saturation vapor pressure. In this study, we developed a simple one-dimensional cloud condensation model including condensation, diffusion and sedimentation of $H_2SO_4$ and $H_2O$ but without detailed microphysics. Our model is able to explain the observed vertical structure of cloud and upper haze mass loading, cloud acidity, $H_2SO_4$, and $H_2O$ vapor, and the mode-2 particle size on Venus. We found that most $H_2SO_4$ is stored in the condensed phase above 48 km, while the partitioning of $H_2O$ between the vapor and clouds is complicated. The cloud cycle is mostly driven by evaporation and condensation of $H_2SO_4$ rather than $H_2O$ and is about seven times stronger than the $H_2SO_4$ photochemical cycle. Most of the condensed $H_2O$ in the upper clouds is evaporated before the falling particles reach the middle clouds. The cloud acidity is affected by the temperature and the condensation-evaporation cycles of both $H_2SO_4$ and $H_2O$. Because of the large chemical production of $H_2SO_4$ vapor and relatively inefficient cloud condensation, the simulated $H_2SO_4$ vapor above 60 km is largely supersaturated by more than two orders of magnitude, which could be tested by future observations.

**Plain Language Summary**

Venus is covered by global thick sulfuric acid clouds. We present a theoretical cloud model without detailed cloud microphysics and successfully explain the vertical structure of sulfuric acid and water vapor, cloud mass loading, and acidity. We found the cloud mass cycle is mostly driven by the evaporation and condensation of $H_2SO_4$ rather than $H_2O$. A large amount of condensed $H_2SO_4$ is evaporated at the cloud base and transported upward. The upward flux is seven times larger than the downward flux from the $H_2SO_4$ source region in the middle atmosphere, indicating strong recycling of $H_2SO_4$ in the lower cloud region. Most of the condensed $H_2O$ in the upper clouds is evaporated before the falling particles reach the middle clouds. Our model predicts that a large excess of $H_2SO_4$ vapor than its saturated value above 60 km due to the large $H_2SO_4$ production rate and relatively inefficient cloud condensation.

**1 Introduction**

Venus is known as a unique terrestrial planet that is covered by thick global sulfuric acid clouds. The clouds significantly regulate the climate on Venus by influencing the atmospheric radiative balance via its high albedo and large infrared opacity (e.g., Arney et al., 2014; Esposito et al., 1983; Seiff et al., 1985; Titov et al., 2018; Tomasko, Smith, et al., 1980). Strong convection in the cloud layers impacts the general circulation and zonal winds on Venus (e.g., Bertaux et al., 2016; Gierasch, 1987).

The aqueous chemistry inside the clouds might explain why the sulfur dioxide—the most abundant sulfur-bearing species on Venus—decreases so rapidly from the lower atmosphere to the cloud top (Rimmer et al., 2021). In the context of habitability, Hallsworth et al. (2021) calculated water activity in the cloud droplets on Venus and found that it was too low to support a habitable environment. Understanding the cloud formation on Venus and the role of clouds on radiation, dynamics, chemistry, and astrobiology in the Venusian climate is thus essential to characterize our sister planet and planetary habitability in general.

The cloud droplets on Venus are mainly composed of concentrated sulfuric acid solution ($H_2SO_4$ and $H_2O$), along with some contaminants like ferric chloride ($FeCl_3$), polysulfide ($S_n$), and ferric sulfate ($Fe_2(SO4)_3$) (Esposito et al., 1983; Krasnopolsky, 1985). The cloud acidity—$H_2SO_4$ weight percent in the droplet—is estimated to be 75%-92% (Arney et al., 2014; Barstow et al., 2012; Cottini et al., 2012; Hansen & Hovenier, 1974; McGouldrick et al., 2021; Pollack et al., 1978). The vertical structure of the clouds on Venus has been revealed by in-situ observations from the cloud particle size spectrometer (LCPS) aboard the Pioneer Venus probes (Knollenberg & Hunten, 1980). The main cloud deck is divided into three layers: the lower cloud from 48 to 50 km, the middle cloud from 50 to 56 km, and the upper cloud from 56 to 70 km. Cloud droplets also show multi-modal particle size distribution that varies with altitude. Mode 1 particles have the largest number density at all three layers with radii less than 0.5 μm. Mode 2 particles are distributed above 48 km and have a mean radius of 1 μm. Mode 3 particles are more likely to be distributed at lower clouds with radii larger than 3.6 μm.

Significant spatial and temporal variations of the Venus cloud properties have also been detected. Decadal fluctuation of the cloud-top sulfur dioxide ($SO_2$) is observed by Pioneer Venus (Esposito et al., 1997) and by Venus Express (Marcq et al., 2012), which might be related to the atmospheric mixing in the clouds (Kopparla et al., 2020; Marcq et al., 2012). The mean cloud top altitude is observed to gradually decline from low latitudes to the polar region by Venus Express (Cottini et al., 2012, 2015; Haus et al., 2013, 2014; Ignatiev et al., 2009; Lee et al., 2012; Zasova et al., 2007). This latitudinal variation is observed to correlate with the mesospheric temperature field (Lee et al., 2012) and the ultraviolet (UV) markings (Ignatiev et al., 2009), and anticorrelated with the cloud-top $H_2O$ mixing ratio (Cottini et al., 2012, 2015). These correlations imply that the Venusian clouds strongly interact with atmospheric dynamics and chemistry.

There are two main differences between the cloud formation on Venus from that on Earth. First, the Venusian clouds are made of the photochemical product ($H_2SO_4$) that is formed in the upper atmosphere. Therefore, the formation of the clouds on Venus is tightly coupled with the sulfur-water chemistry (Bierson & Zhang, 2020; Krasnopolsky, 2012; Mills, 1998; Mills & Allen, 2007; Shao et al., 2020; Yung & Demore, 1982; Zhang, Liang, et al., 2012). The main sulfur-bearing gas sulfur dioxide and main hydrogen-carrier $H_2O$ vapor are transported from the lower atmosphere to the upper atmosphere. At the cloud top, $SO_2$ is photolyzed and produces sulfur trioxide ($SO_3$), which combines with $H_2O$ to produce $H_2SO_4$ (Bierson & Zhang 2020; Krasnopolsky, 2012; Krasnopolsky & Parshev, 1981; Mills, 1998; Shao et al., 2020; Winick & Stewart, 1980; Yung & Demore, 1982). The $H_2SO_4$ vapor subsequently condenses with $H_2O$ to form globally covered clouds on Venus. Second, the Venusian cloud is formed from binary condensation of $H_2SO_4$ and $H_2O$. $H_2SO_4$ is produced in the upper cloud region and water is transported upward from the cloud bottom. The saturation vapor pressure (SVP) of $H_2SO_4$ and $H_2O$ significantly varies with the droplet acidity (Giauque et al., 1960; Zeleznik, 1991), leading to complex interactions between the vapor and condensed phases of $H_2SO_4$ and $H_2O$. Moreover, it was recently proposed that hydroxide salts could contribute a significant fraction of the Venusian clouds (Rimmer et al., 2021). The possible aqueous chemistry and its effect on the cloud acidity further complicate the understanding of the system.

Several models have been established to understand the formation of Venus clouds. Before the observation of Pioneer Venus, Rossow & Gierasch (1977) and Rossow (1977) developed a simple Venusian cloud model using the moment method with coagulation physics, which well predicted the particle distribution later observed by Pioneer Venus (Knollenberg & Hunten, 1980). But their model did not treat the condensation processes using supersaturation and the interaction between the saturation vapor pressure of $H_2SO_4$ and $H_2O$ and the cloud acidity. Later studies adopted more complicated microphysical approaches including particle nucleation, condensation, coagulation, evaporation to explain the properties of the main clouds below 70 km (Hashimoto & Abe, 2001; Imamura & Hashimoto, 2001; James et al., 1997; McGouldrick, 2007; McGouldrick & Toon, 2007; Toon et al., 1982). Gao et al. (2014) and Parkinson et al. (2015) further extended the microphysical approach to the upper mesosphere to explain the upper hazes from the Pioneer Venus (Kawabata et al., 1980) and Venus Express (Wilquet et al., 2009, 2012). Some microphysical models treated the binary condensation of $H_2SO_4$ and $H_2O$ interactively with the cloud acidity (Imamura & Hashimoto, 2001; James et al., 1997; McGouldrick & Toon, 2007) but others did not (Gao et al. 2014, Parkinson et al. 2015). On the other hand, Krasnopolsky & Pollack (1994) developed a model (updated by Krasnopolsky, 2015) without microphysics and explained the cloud acidity and vapor distributions by assuming local thermodynamic equilibrium (LTE) between the droplets and vapor. But their model did not calculate and explain the cloud particle distribution and the cloud mass loading. As we will show later, LTE is not a valid assumption in the source region of the $H_2SO_4$ gas. Our finding might also impact the gas chemical simulation of Venus because some previous models assumed that the $H_2SO_4$ vapor follows LTE (e.g., Zhang et al. 2010, 2012).

The goal of our study is to provide a one-dimensional (1D) cloud model on Venus that does not include detailed microphysics but is able to simultaneously explain four critical observations: the vertical distributions of $H_2SO_4$ vapor and $H_2O$ vapor, as well as that of the cloud acidity and mass loading including the main clouds and the upper hazes {\footnote: The mass loading from Wilquet et al. (2009) is not from a direct observation, but rather an estimate based on the assumed size distribution. Because the size distribution is consistent with a previously published microphysical model in Gao et al. (2014), we think it is still meaningful to compare our model results with the estimated upper haze mass based on Wilquet et al. (2009) data}. Although our model does not simulate the particle size distribution, we can still simulate a particle size profile that is roughly consistent with the mass-averaged size based on observations from the Pioneer Venus (Knollenberg & Hunten, 1980). With this simple cloud model, the mechanisms governing the distributions of $H_2SO_4$ and $H_2O$ in both vapor and condensed phases on Venus can be clearly diagnosed and described. The mass flows and important chemical and cloud cycles of $H_2SO_4$ and $H_2O$ are shown in Figure 1. The details will be discussed later with our model results.

This paper is organized as follows. In section 2 we introduce our model and equations. In section 3 we show the results of the nominal simulation and data-model comparison. We will also investigate the underlying mechanisms behind the $H_2SO_4$ and $H_2O$ cycles. In section 4 we demonstrate the sensitivity of the model results to critical input parameters. The comparisons with other models are described in section 5. We conclude this study in section 6.

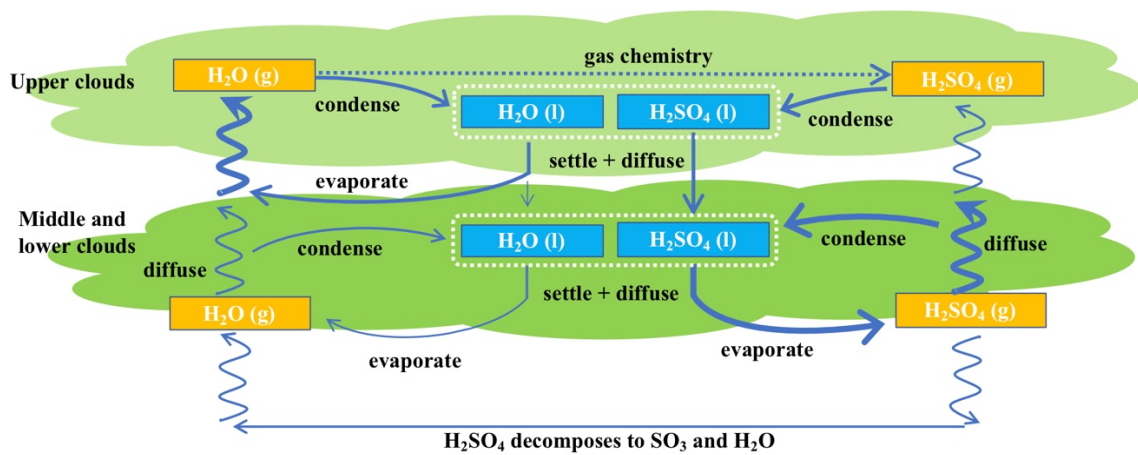

**Figure 1.** Schematic cartoon of chemical and cloud cycles of $H_2SO_4$ and $H_2O$ in the atmosphere of Venus. The line thickness indicates the relative magnitude of the mass fluxes and the cycle strength. See details in section 3. Note that the condensed $H_2SO_4$ and $H_2O$ are mixed in the droplets as denoted by the dashed line boxes.

**2 Model description**

Our model solves one-dimensional continuity equations for both gas and condensed species with important source and sink terms ($S$) including the eddy diffusion of gas ($S_{diff}^g$) and condensed ($S_{diff}^c$) species, net gas chemical production/loss ($S_{chem}^g$), condensation/evaporation ($S_{cond}$), and particle sedimentation ($S_{sed}^c$). For each species, the continuity equation is written as

$$\frac{\partial n}{\partial t} = S_{diff} + S_{chem} + S_{cond} + S_{sed}, \tag{1}$$

where $n$ is the abundance of the species—the number of molecules in each phase within an atmospheric volume. We use subscript 1 and 2 for $H_2SO_4$ and $H_2O$, respectively, and superscript $g$ and $c$ for the gas and condensed phase, respectively. Our model includes four species: $H_2SO_4$ vapor ($n_1^g$), $H_2O$ vapor ($n_2^g$), condensed $H_2SO_4$ ($n_1^c$), and condensed $H_2O$ ($n_2^c$). We discuss the important source and sink terms below.

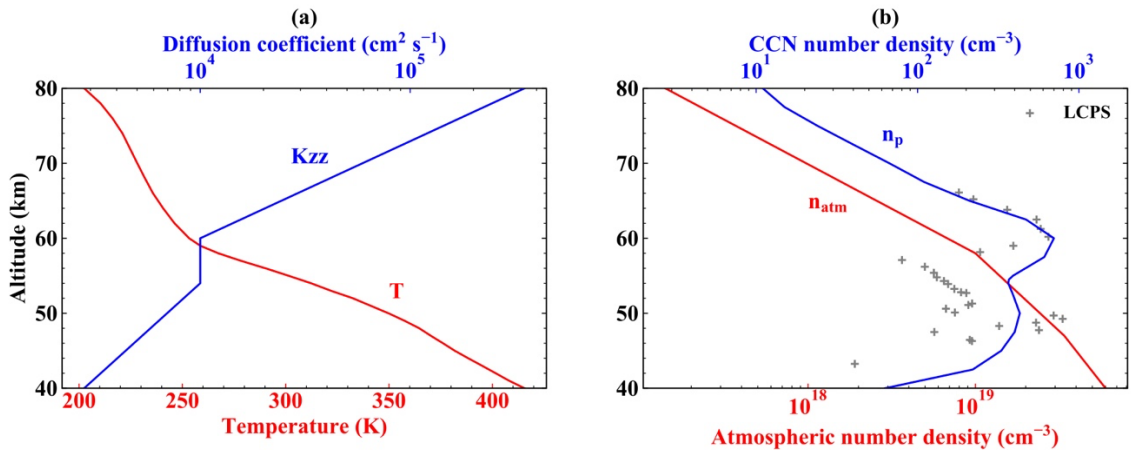

**Figure 2.** Prescribed atmospheric properties in the model: (**a**) the vertical distribution of the atmospheric temperature (red) and the eddy diffusion coefficient (blue). The temperature profile is adopted from the observation of Pioneer Venus in 45° latitudes (Seiff et al., 1985). The diffusion coefficient above 45 km is adopted from Krasnopolsky (2012, 2015) and it is assumed to exponentially reduce to 2800 cm$^2$ s$^{-1}$ at 40 km; (**b**) the vertical distribution of the atmospheric number density (red) and the CCN number density (blue). The atmospheric number density profile is adopted from Krasnopolsky (2012). The CCN number density profile is adopted from Gao et al. (2014). The observation of CCN number density (grey pluses) is from LCPS (Knollenberg & Hunten, 1980).

### 2.1 Condensation and evaporation

The condensation (evaporation) occurs when the vapor partial pressure exceeds (falls short of) the SVP. The efficiency of this process depends on multiple parameters such as supersaturation, molecular diffusion, temperature, particle size, and surface tension. In our simple model, we neglect the particle size distribution and parameterize the particle growth rate following Seinfeld & Pandis (2016, p.538-540):

$$\frac{dD_p}{dt} = \frac{4D_i M_i}{R_g T D_p \rho_p}(P_i - P_i^{svp})f_{m_i}, \tag{2}$$

where $D_p$ is the diameter of the particle, $R_g$ is the gas constant, $T$ is the atmospheric temperature, $D_i$ is the molecular diffusion coefficient, $M_i$ is the relative molecular weight of species $i$, $P_i$ is the partial pressure of species $i$ far from the particle surface and $P_i^{svp}$ is the SVP of species $i$ over the particle surface. $\rho_p = 1.8$ g cm$^{-3}$ is the density of the droplets, which we assumed constant in this study. We note that the density of sulfuric acid droplets changes with temperature and acidity (Myhre et al., 1998) but our test shows that including the acidity and temperature dependence of the droplet density does not alter our results. The temperature profile is adopted from the observation of Pioneer Venus at 45° latitude (Seiff et al., 1985), shown in Figure 2a. The $D_i$ is calculated following Hamill et al. (1977):

$$D_i = \frac{\lambda_i}{3}\left(\frac{8kTN_a}{\pi M_i}\right)^{\frac{1}{2}} \tag{3}$$

where $\lambda_i$ is mean free path of species $i$, $N_a$ is the Avogadro constant, $k$ is the Boltzmann constant. $\lambda_i$ is calculated following Hamill et al. (1977):

$$\lambda_i = \left(\pi n_{atm}\left(\frac{(d_i + d_{atm})}{2}\right)^2 \left(\frac{M_{atm}}{M_i + M_{atm}}\right)^{\frac{1}{2}}\right)^{-1} \tag{4}$$

where $d_i$ is the molecular diameter of species $i$, $d_{atm}$ is the molecular diameter of atmospheric major component (carbon dioxide), $M_{atm} = 44$ g/mol is the relative molecular weight of atmospheric major component. We take $d_1 = 4.4 \times 10^{-8}$ cm (assumed in Hanson & Eisele, 2000), $d_2 = 2.65 \times 10^{-8}$ cm (Matteucci et al., 2006), $d_{atm} = 3.3 \times 10^{-8}$ cm (Mehio et al., 2014). The flux-matching factor $f_m$ is adopted from Zhang, Pandis, et al. (2012):

$$f_{m_i} = \frac{1}{1 + \frac{2\lambda_i}{D_p \alpha}}, \tag{5}$$

where $\alpha = 1$ is the accommodation coefficient (Seinfeld & Pandis, 2016, p.500-502).

In our model, the droplets grow via vapor condensation onto small cloud condensation nuclei (CCN, $r_s = 4 \times 10^{-3}$ µm) that are assumed to be nucleated already. Without detailed microphysics, our model does not solve for CCN. Instead, the vertical profile of CCN number density ($n_p$, Figure 2b) is prescribed as that in Gao et al. (2014), which matches the LCPS observations (Knollenberg & Hunten, 1980) and the upper haze estimations (Wilquet et al., 2009). We also did a sensitivity test to explore the impact of the adopted CCN profile on our simulation results in section 4.4.

Given the particle number density (CCN number density) and the molecular ratio of the condensed species $H_2SO_4$ and $H_2O$, the particle size growth rate (equation 2) can be directly converted to the condensate/evaporation rate $S_{cond}$ in equation (1). Assuming that the CCN mass is negligible compared with the droplet mass, the particle diameter $D_p$ is related to the droplet mass by

$$D_p = 2\left(\frac{3(n_1^c M_1 + n_2^c M_2)}{4\pi N_a n_p \rho_p} + r_s^3\right)^{\frac{1}{3}}. \tag{6}$$

The molecular abundance of each condensate in the droplets is:

$$n_1^c = \frac{\frac{\pi}{6} D_p^3 \rho_p n_p N_a}{M_1 + M_2 m} \tag{7}$$

and

$$n_2^c = m n_1^c = \frac{\frac{\pi}{6} D_p^3 \rho_p n_p N_a}{\frac{M_1}{m} + M_2}, \tag{8}$$

where $m = n_2^c/n_1^c$ is the $H_2O/H_2SO_4$ molecular ratio in the droplet. Putting equation (2) - (8) together and using the ideal gas law, we can obtain the condensation/evaporation rate equation for $H_2SO_4$:

$$S_{cond} = \frac{\partial n_1^c}{\partial t} = \frac{\pi \rho_p n_p N_a}{6(M_1 + M_2 m)} \cdot 3D_p^2 \frac{\partial D_p}{\partial t} = \frac{2\pi n_p D_1 M_1 f_{m_i} n_{atm} D_p}{(M_1 + M_2 m)} \left(q_1^g - q_1^{svp}\right), \tag{9}$$

and that for $H_2O$:

$$S_{cond} = \frac{\partial n_2^c}{\partial t} = \frac{\pi \rho_p n_p N_a}{6\left(\frac{M_1}{m} + M_2\right)} \cdot 3D_p^2 \frac{\partial D_p}{\partial t} = \frac{2\pi n_p D_2 M_2 f_{m_i} n_{atm} D_p}{\left(\frac{M_1}{m} + M_2\right)} \left(q_2^g - q_2^{svp}\right). \tag{10}$$

Here $q_i^{svp}$ is the saturation vapor mixing ratios (SVMR) of species $i$. The rate is proportional to the CCN number density $n_p$, droplet diameter $D_p$ and the supersaturation $P_i - P_i^{svp}$. We define $q_i^g = n_i^g/n_{atm}$ and $q_i^c = n_i^c/n_{atm}$ as the volume mixing ratios (molar fractions) of the gas and condensed phase $i$, respectively, with respect to the total atmospheric number density $n_{atm}$. The vertical profile of $n_{atm}$ (Figure 2b) is adopted from Krasnopolsky (2012).

The SVPs of both $H_2SO_4$ and $H_2O$ are affected by the $H_2O/H_2SO_4$ molecular ratio $m$ in the cloud droplets, measured by:

$$\ln P_i^{svp} = \ln P_{i,p}^{svp} + \frac{\mu_i - \mu_i^0}{R_g T} + \frac{4\sigma M_i}{R_g T \rho_p D_p}, \tag{11}$$

where $P_{i,p}^{svp}$ is the SVP (in units of bar) above the pure condensed species $i$, $\mu_i - \mu_i^0$ is the difference of the chemical potential of species $i$ between the solution and its pure condensate. This chemical potential difference depends on the $H_2O/H_2SO_4$ molecular ratio in the condensed droplets and the temperature (Zeleznik, 1991). The third term on the right-hand side comes from the Kelvin effect due to the curvature of the droplet surface (see Seinfeld & Pandis, 2016, p.419-423), where $\sigma = 71.11$ erg cm$^{-2}$ is the surface tension of sulfuric acid (Myhre et al., 1998, 80% sulfuric acid, 273 K), which we assumed constant in this study. We note that the surface tension of the sulfuric acid droplet changes with temperature and acidity (Vehkamäki et al., 2003) but our test shows that this effect is negligible in our study. The value of $P_{1,p}^{svp}$ is from Kulmala & Laaksonen (1990):

$$\ln P_{1,p}^{svp} = 16.259 - \frac{10156}{T_0} + 10156\left[-\frac{1}{T} + \frac{1}{T_0} + \frac{0.38}{T_c - T_0} \times \left(1 + \ln\frac{T_0}{T} - \frac{T_0}{T}\right)\right], \quad (12)$$

where $T_0 = 360$ K is a reference temperature, $T_c = 905$ K is the critical temperature. For $H_2O$, we adopt the $P_{2,p}^{svp}$ value from Tabazadeh et al. (1997):

$$\ln P_{2,p}^{svp} = \exp\left(18.4524 - \frac{3505.2}{T} - \frac{330918.6}{T^2} + \frac{12725068.3}{T^3}\right). \quad (13)$$

The SVMRs of both species as functions of altitude and the cloud acidity are shown in Figure 3. The variation of the cloud acidity has a great impact on the SVMRs of both species, especially for $H_2O$ in lower altitudes. Increasing the cloud acidity from 75% to 98% at 50 km leads to the decrease of $H_2O$ SVMR by four orders of magnitude and the increase of $H_2SO_4$ SVMR by two orders of magnitude. Thus, it is essential to self-consistently model the cloud acidity and the vapor abundances.

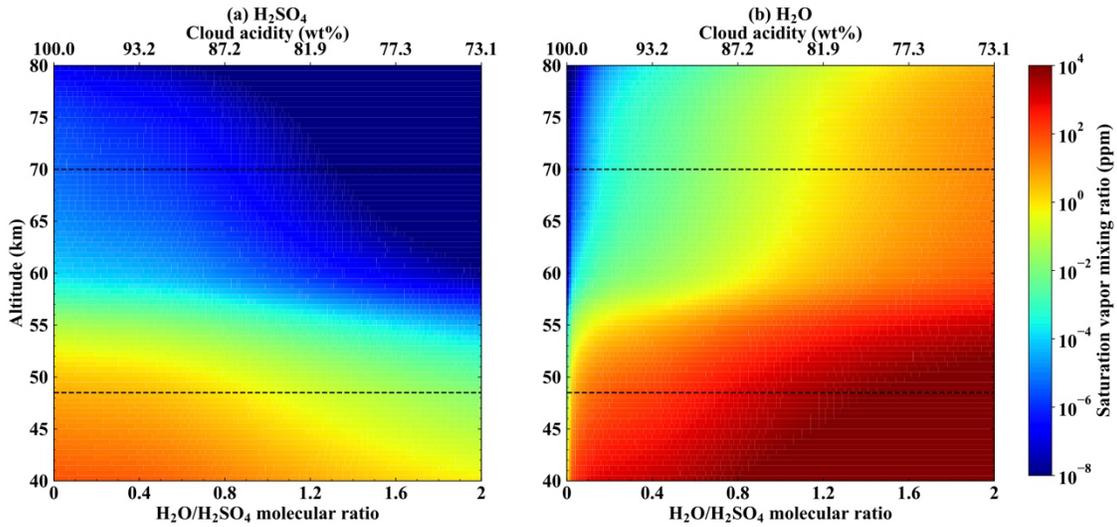

**Figure 3.** The saturation vapor mixing ratios of $H_2SO_4$ (**a**) and $H_2O$ (**b**) as functions of the altitude and the cloud acidity. The droplet $H_2O/H_2SO_4$ molecular ratio is also shown in the bottom x-axis. We did not include the Kelvin effect in this plot because it depends on the particle size. We choose an acidity range to cover the estimated range from observations (75% ~ 92%, Arney et al., 2014; Barstow et al., 2012; Cottini et al., 2012; Hansen & Hovenier, 1974; McGouldrick et al., 2021; Pollack et al., 1978) and the simulated range from previous models (e.g., 75% ~ 98%, Krasnopolsky, 2015). The saturation vapor mixing ratios are calculated by equation (11) - (13). The black dashed lines represent the approximate cloud top and the cloud base, which are based on the observations of Venus Express (Barstow et al., 2012; Cottini et al., 2012; Ignatiev et al., 2009; Lee et al., 2012; McGouldrick et al., 2021) and Pioneer Venus (Knollenberg & Hunten, 1980), respectively.

2.2 Eddy Diffusion

The gas species are assumed to be mixed by turbulent eddies. Molecular diffusion is ignored in this study since our upper boundary (80 km) is far below the homopause (~125 km). The eddy diffusion term (divergence/convergence) of the gas-phase species is

$$S_{diff_i}^g = -\frac{\partial}{\partial z}\left(-K_{zz} n_{atm} \frac{\partial q_i^g}{\partial z}\right), \qquad (14)$$

where $z$ is altitude, $K_{zz}$ is the eddy diffusion coefficient shown in Figure 2a. Both the eddy diffusion coefficient and the photochemical production rate of sulfuric acid vapor are adopted from a state-of-the-art photochemical model in Krasnopolsky (2012), which spans from 47 km to 112 km. This profile is mainly constrained by the latest observations and is consistent with the eddy diffusion coefficient retrieved from the aerosol data (Lane & Opstbaum, 1983) and the upper limit is inferred from scintillations of the radio signal (Woo & Ishimaru, 1981). Below the cloud base, Krasnopolsky & Pollack (1994) inferred smaller eddy diffusivity in the sub-cloud atmosphere than in the cloud layer, according to the large wind shear at the cloud bottom (Kerzhanovich & Marov, 1983) and the lower static stability in the cloud layer (Linkin et al., 1986; Pollack et al., 1980). In this study, we followed Krasnopolsky (2015) and assumed that the eddy diffusion coefficient drops exponentially below 54 km to 2800 cm$^2$ s$^{-1}$ at 40 km. But we also noted that the eddy diffusivity is uncertain inside the clouds. Some previous studies also adopted a profile with a local maximum in the middle cloud (Gao et al., 2014; Imamura & Hashimoto, 2001; James et al., 1997; McGouldrick & Toon, 2007). McGouldrick & Toon (2007) also estimated the eddy diffusion as a function of the evolving static stability profile. In section 4.2, we tested a $K_{zz}$ profile with a local maximum inside the cloud layer from Imamura & Hashimoto (2001) and investigate its impact on our model results.

The condensed species are assumed to diffuse with the particles in the cloud layer. The eddy diffusion term of condensed species $i$ is

$$S^c_{diff_i} = -\frac{\partial}{\partial z}\left(-K_{zz}n_{atm}\frac{\partial \frac{M_p}{n_{atm}}}{\partial z}\right)\left(\frac{M_i q^c_i}{\sum_i(M_i q^c_i)}\right)\frac{N_a}{M_i}, \qquad (15)$$

where $M_p$ is the cloud mass loading:

$$M_p = (M_1 q^c_1 + M_2 q^c_2)\frac{n_{atm}}{N_a}. \qquad (16)$$

2.3 Gas Chemical Production and Loss

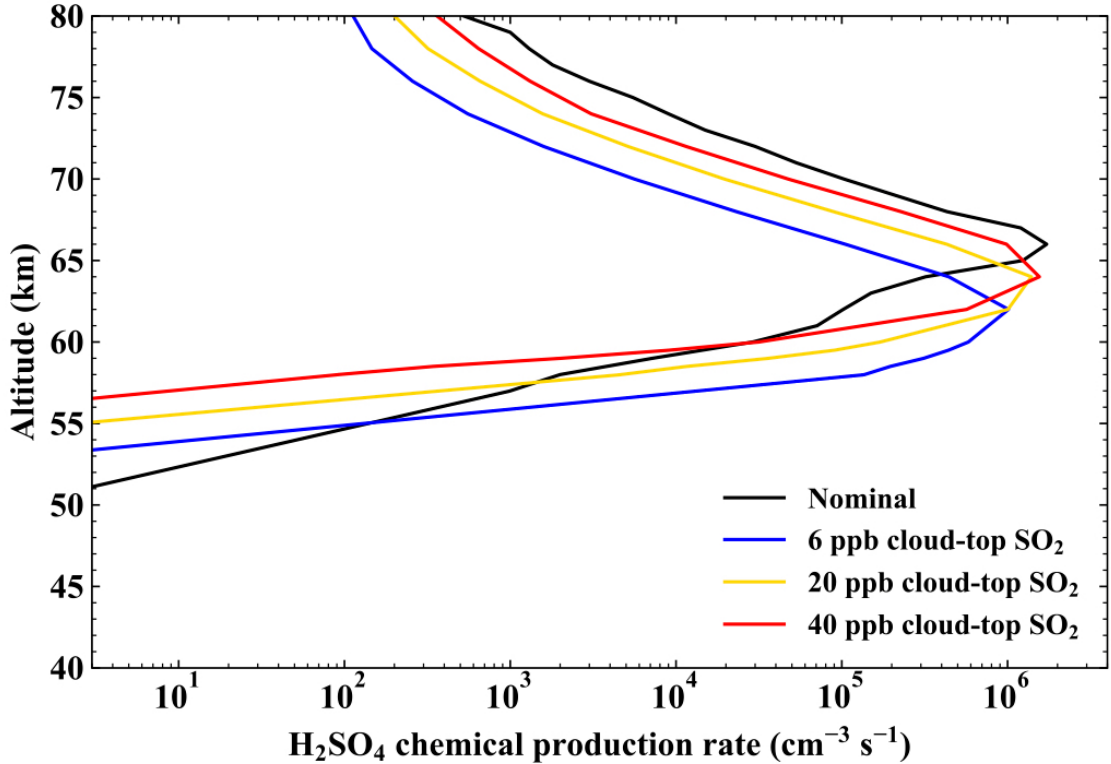

**Figure 4.** The prescribed net chemical production rates of $H_2SO_4$ in our model. The nominal profile (black) is adopted from Krasnopolsky (2012). Three other $H_2SO_4$ production rate profiles from Shao et al. (2020) are used for the sensitivity tests in section 4.1. These profiles correspond to a cloud-top (70 km) $SO_2$ mixing ratios of 6 ppb (blue), 20 ppb (yellow), and 40 ppb (red), respectively. The $SO_2$ mixing ratio at the cloud top in Krasnopolsky (2012) is about 100 ppb. The $H_2O$ chemical loss rate is equivalent to this $H_2SO_4$ production rate.

We did not solve chemistry in this model. Instead, we prescribed the chemical net production rate ($S_{chem}$) of $H_2SO_4$ and chemical net loss rate of $H_2O$ in the mesosphere. The reaction $SO_3+2H_2O=H_2SO_4+H_2O$ is the main source of $H_2SO_4$ and the main sink of $H_2O$ (Bierson & Zhang 2020; Krasnopolsky, 2012; Krasnopolsky & Parshev, 1981; Mills, 1998; Shao et al., 2020; Winick & Stewart, 1980; Yung & Demore, 1982) and thereby the two rates are assumed the same in our model. 1D photochemical models (Krasnopolsky, 2012; Mills, 1998; Yung & Demore, 1982; Zhang, Liang, et al., 2012) demonstrate that the photodissociation of $H_2O$ is negligible in the cloud region. $SO_3$ production is mainly controlled by the chemical reaction between $SO_2$ and O, and almost all $SO_3$ is converted into $H_2SO_4$ (Krasnopolsky, 2012; Mills, 1998; Shao et al., 2020; Yung & Demore, 1982). According to previous photochemical models (Krasnopolsky, 2012; Mills, 1998; Shao et al., 2020; Winick and Stewart, 1980; Yung & Demore, 1982), the $H_2SO_4$ production rate is primarily limited by the $SO_3$ abundance because $H_2O$ is sufficient. The cloud formation does not significantly alter the abundance of $SO_2$ and O. We also note that different chemical models yield different $H_2SO_4$ production profiles. In early models, $H_2SO_4$ gas was mainly produced in a narrow layer by photochemical processes with a peak reaction rate at 62 km (Krasnopolsky & Parshev, 1981; Mills, 1998; Yung & Demore, 1982). Later photochemical models (Krasnopolsky, 2012; Shao et al., 2020; Zhang, Liang, et al., 2012) show higher-altitude $H_2SO_4$ production peaks, near 66 km. Furthermore, $SO_2$ have been observed to exhibit long-term variations (Encrenaz et al., 2016, 2019, 2020; Esposito, 1984; Esposito et al., 1988, 1997; Marcq et al., 2012; Na et al., 1990; Vandaele et al., 2017). For our nominal case, we adopt the $H_2SO_4$ production rate profile from Krasnopolsky (2012), as shown in Figure 4. The column-integrated $H_2SO_4$ production rate is about $5.7 \times 10^{11}$ cm$^{-2}$ s$^{-1}$, which is consistent with $6 \times 10^{11}$ cm$^{-2}$ s$^{-1}$ in Gao et al. (2014) and smaller than $10^{12}$ cm$^{-2}$ s$^{-1}$ in Imamura & Hashimoto (2001). As we will show, the shape of the production rate profile of $H_2SO_4$ has a significant impact on the supersaturation of $H_2SO_4$ vapor in the middle atmosphere. We will discuss the effect of different $H_2SO_4$ production profiles on our results in sections 3.5 and 4.1.

The thermal decomposition of $H_2SO_4$ to $SO_3$ and $H_2O$ appears significant below 38 km from the observations of Magellan (Jenkins et al., 1994). Since we do not solve chemistry in this model, we follow the previous microphysical models (Gao et al., 2014; Imamura & Hashimoto, 2001; James et al., 1997; McGouldrick & Toon, 2007) and neglect the $H_2SO_4$ thermal decomposition above 40 km in our model.

2.4 Particle Sedimentation

In our model, the droplets are assumed to settle down with the Stokes velocity:

$$v = -\frac{2}{9} \frac{g\rho_p D_p^2}{4\eta} C_c, \qquad (17)$$

where $g = 870$ cm s$^{-2}$ is the gravitational acceleration on Venus. The dynamic viscosity of $CO_2$ gas $\eta$ varies with temperature and the data is from Fenghour et al. (1998). The Cunningham slip correction factor $C_c$ is calculated following Seinfeld & Pandis (2016, p. 371-372):

$$C_c = 1 + \frac{2\lambda}{D_p}\left(1.257 + 0.4\, e^{\frac{-1.1 D_p}{2\lambda}}\right), \tag{18}$$

where $\lambda$ is the mean free path. Thus the sedimentation term $S_{sed}^c$ in equation (1) is:

$$S_{sed}^c = -\frac{\partial}{\partial z}(v n_{atm} q_i^c). \tag{19}$$

2.5 Boundary conditions and numerical setup

**Table 1.** *Model Parameters and Boundary Conditions.*

| Parameter | Nominal value |
| --- | --- |
| $T$, $K_{zz}$ profile | Figure 2a |
| $n_{atm}$, $n_p$ profile | Figure 2b |
| $H_2SO_4$ production profile | Figure 4 |
| $D_i$ | Equation (3) |
| $\lambda_i$ | Equation (4) |
| $\eta$ | Fenghour et al. (1998) |
| $M_1$ | 98 g mol$^{-1}$ |
| $M_2$ | 18 g mol$^{-1}$ |
| $M_{atm}$ | 44 g mol$^{-1}$ |
| $\rho_p$ | 1.8 g cm$^{-3}$ |
| $R$ | 8.314×10$^7$ g cm$^2$ s$^{-2}$ mol$^{-1}$ K$^{-1}$ |
| $k$ | 1.381×10$^{-16}$ g cm$^2$ s$^{-2}$ K$^{-1}$ |
| $r_s$ | 4×10$^{-3}$ μm |
| $g$ | 870 cm s$^{-2}$ |
| $\alpha$ | 1 |
| $\sigma$ | 71.11 erg cm$^{-2}$ |
| $d_1$ | 4.4×10$^{-8}$ cm |
| $d_2$ | 2.65×10$^{-8}$ cm |
| $d_{atm}$ | 3.3×10$^{-8}$ cm |
| $T_0$ | 360 K |
| $T_c$ | 905 K |
| | |
| Upper boundary (80 km) | |
| $\quad$ H$_2$SO$_4$ vapor | $\partial q_1^g / \partial z = 0$ |
| $\quad$ H$_2$O vapor | $\partial q_2^g / \partial z = 0$ |
| $\quad$ Condensed H$_2$SO$_4$ | $\partial q_1^c / \partial z = 0$ |
| $\quad$ Condensed H$_2$O | $\partial q_2^c / \partial z = 0$ |
| Lower boundary (40 km) | |
| $\quad$ H$_2$SO$_4$ vapor | $q_1^g = 3$ ppm (parts per million) |
| H₂O vapor | $q_2^g = 27$ ppm |
| Condensed H₂SO₄ | $q_1^c = 0$ ppm |
| Condensed H₂O | $q_2^c = 0$ ppm |

This model extends from 40 to 80 km, covering the altitude range of the cloud layer and part of lower and upper haze. The vertical resolution is 0.5 km. We solved the equation set using a semi-implicit scheme. For gas, the diffusion and condensation terms are treated implicitly, and the sedimentation and chemical production/loss terms are treated explicitly. Because of the interaction of the cloud mass and the cloud acidity, we solve the condensates explicitly. The detailed numerical scheme is shown in Appendix A. A typical simulation reaches the steady state after integrating the model for 800 Earth days, which takes about 15-30 minutes on a single CPU core. The boundary conditions and the model parameters are summarized in Table 1.

Previous photochemical models (Krasnopolsky, 2012; Krasnopolsky & Parshev, 1981; Mills, 1998; Yung & Demore, 1982; Zhang, Liang, et al., 2012) found the $H_2SO_4$ production rate rapidly decreases as altitude increases above the production peak at 66 km and is negligible at 80 km. We also neglected the hydrogen escape and loss in the upper boundary. Therefore, we adopt the zero flux boundary conditions for all the species at the upper boundary. These boundary conditions are consistent with Imamura & Hashimoto (2001). $H_2O$ is the main reservoir of hydrogen in the cloud layers with the observed mixing ratio of 30±10 ppm below 40 km (Bézard et al., 2009; Chamberlain et al., 2013; Marcq et al., 2008). We fixed the total hydrogen ($H_2$) mixing ratio is 30 ppm at the lower boundary with 3 ppm of $H_2SO_4$ vapor (same as Gao et al., 2014) and 27 ppm of $H_2O$ vapor. These mixing ratios are consistent with observations from Venus Express (Marcq et al., 2008; Oschlisniok et al., 2012, 2021) and ground-based spectroscopy (de Bergh et al., 1991, 1995; Pollack et al., 1993; Arney et al., 2014). The condensed $H_2SO_4$ and $H_2O$ are set as zero mixing ratios at the lower boundary which is far below the cloud base as observed by the Pioneer Venus (Knollenberg & Hunten, 1980), Magellan Mission (Kolodner & Steffes, 1998), Venus Express (Barstow et al., 2012; McGouldrick et al., 2021), and ground-based spectroscopy (Pollack et al., 1993).

**3 Simulation results**

In the following, we presented the simulated vertical distributions of gas species, cloud mass loading, cloud acidity, and particle diameter and sedimentation velocity and compare them with observations. In addition, we investigated the $H_2SO_4$ cycle and the water cycle in detail to understand the mechanisms controlling the vertical profiles of the vapor and cloud properties. Both cycles have been summarized in a cartoon in Figure 1 to illustrate the important mass flows.

3.1 $H_2SO_4$

The partitioning of $H_2SO_4$ in the vapor and condensed phase is illustrated in Figure 5a. The distribution of the $H_2SO_4$ vapor mixing ratio is consistent with the observations of Magellan Mission (Kolodner & Steffes, 1998), Akatsuki Mission (Imamura et al., 2017), Venus Express (Oschlisniok et al., 2012, 2021). The simulated $H_2SO_4$ at 80 km is 0.02 ppb, far below the 85-km upper limit of 3 ppb detected by ground-based spectroscopy (Sandor et al., 2012). The $H_2SO_4$ vapor mixing ratio increases from our fixed lower boundary value of 3 ppm to the peak of 5 ppm at the cloud base. Then it rapidly decreases with increasing altitude, following the SVMR. Another inversion exists above 60 km, forming a supersaturated $H_2SO_4$ layer with a peak of $10^{-4}$ ppm at 66 km. The condensed $H_2SO_4$ is much more abundant than the $H_2SO_4$ vapor above 50 km. For example, the condensable vapor mass is about $6\times10^{-6}$ mg m$^{-3}$ compared to the condensed mass of 1 mg m$^{-3}$ at 66 km. The vertical profile of the condensed $H_2SO_4$ shows two peaks of 2.5 and 5 ppm at the cloud base and around 65 km, respectively. The total $H_2SO_4$ (gas and condensate) mixing ratio has the same peak as condensed $H_2SO_4$ at around 65 km and a larger peak of 7 ppm at the cloud base, then it follows the vapor below the clouds.

As shown in Figure 1, the $H_2SO_4$ cycle in the Venus atmosphere is composed of a weaker chemical cycle (in term of the fluxes and rates) coupling the lower and upper atmosphere and a stronger cloud condensation-evaporation cycle inside the cloud layers between 48 and 56 km. The $H_2SO_4$ vapor is formed near 66 km and quickly condensed into the droplets. Almost all $H_2SO_4$ is transported downward in the condensed phase instead of the vapor phase. Once the droplets reach the cloud base at about 48 km, most of the evaporated $H_2SO_4$ vapor is transported upward and re-condensed into the clouds until 56 km. The strength of this strong cloud condensation-evaporation cycle is determined by droplet growth, sedimentation, and diffusion as well as vapor diffusion, and consequently affects the cloud mass loading and cloud acidity. It is this strong cloud cycle that forms most of the thick middle and lower clouds on Venus in our model. The rest of the evaporated vapor is transported downward below the cloud base. Because we did not include chemistry in the lower atmosphere, we fixed the lower boundary value of $H_2SO_4$ as 3 ppm at 40 km to force the downward flux of $H_2SO_4$ below the cloud base. On Venus, $H_2SO_4$ will be thermally decomposed back into $SO_3$ and $H_2O$ in the lower atmosphere according to previous chemical models (Krasnopolsky, 2007) and observations (Jenkins et al., 1994). The $SO_3$ is further converted to $SO_2$ and diffused upward with water to produce the $H_2SO_4$ vapor in the upper atmosphere via photochemistry and close the sulfur chemical cycle.

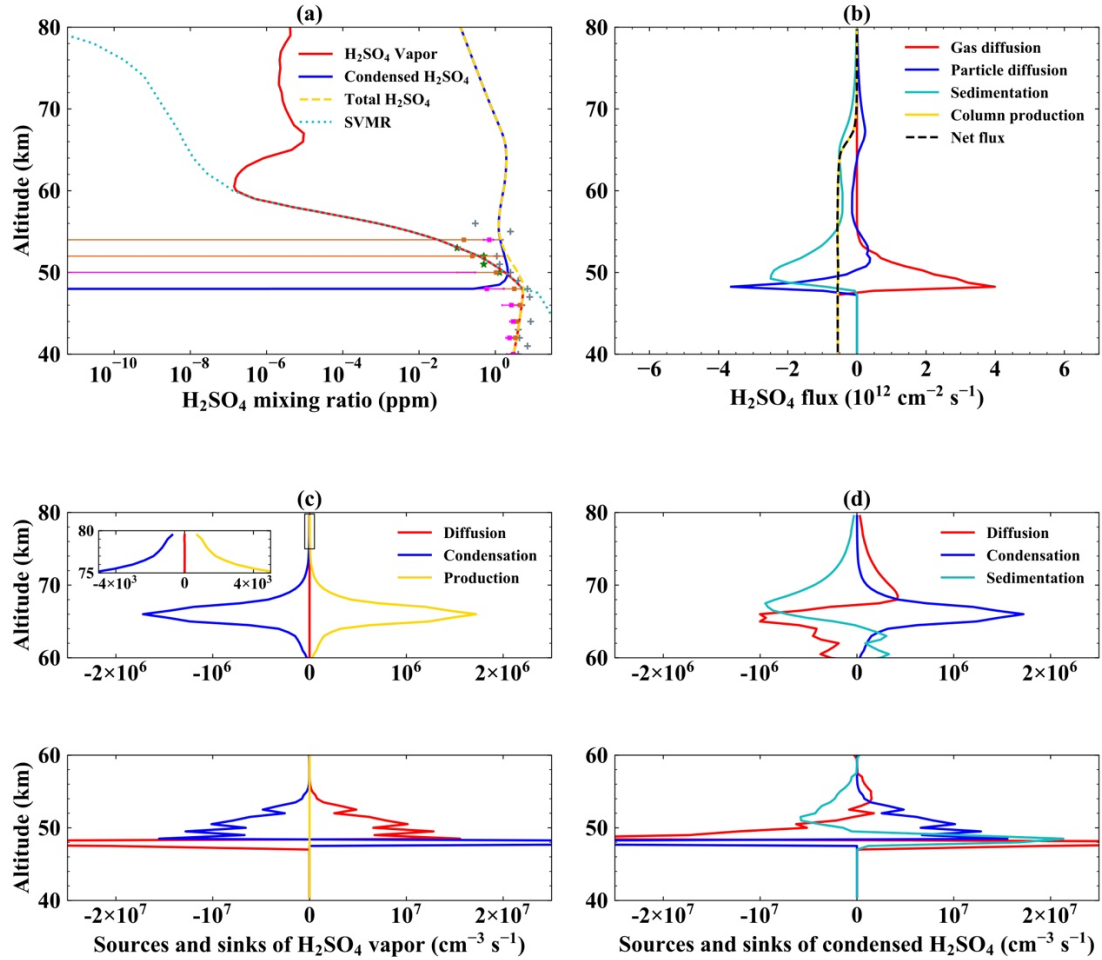

**Figure 5.** The $H_2SO_4$ simulation results in the nominal case. (**a**) the volume mixing ratios of $H_2SO_4$ vapor (red solid line) and the condensed $H_2SO_4$ (blue solid line). The total $H_2SO_4$ mixing ratio is shown as the yellow dashed line. The $H_2SO_4$ SVMR is shown as the cyan dotted line. The observations of the $H_2SO_4$ vapor mixing ratio are from Magellan Mission (chocolate bar, Kolodner & Steffes, 1998), Venus Express (green star, Oschlisniok et al., 2012; pink bar, Oschlisniok et al., 2021), and Akatsuki Mission (grey cross, Imamura et al., 2017). (**b**) the $H_2SO_4$ fluxes by the gas diffusion (red), the particle diffusion (dark blue), and sedimentation (cyan). The positive value means that the flux is upward. The net flux is the black dashed line. The column-integrated production rate of $H_2SO_4$ (yellow) is $5.7\times10^{11}$ cm$^{-2}$ s$^{-1}$ (Krasnopolsky, 2012). (**c**) the source and sink terms of $H_2SO_4$ vapor including diffusion (red), condensation or evaporation (blue), and chemical production (yellow). The region 75-80 km is plotted in a subplot at the upper left corner. (**d**) the source and sink terms of $H_2SO_4$ in condensed phases including particle diffusion (red), condensation or evaporation (blue), and particle sedimentation (cyan).

The entire $H_2SO_4$ cycle can be understood using the upward (positive) and downward (negative) $H_2SO_4$ fluxes (Figure 5b) and the associated sources and sinks (Figure 5c and 5d). The $H_2SO_4$ cycle in the lower cloud of Venus is mostly dominated by particle diffusion and sedimentation, and gas diffusion and re-condensation. At the cloud base, the total downward flux (particle diffusion and sedimentation) reaches $-4.5 \times 10^{12}$ $cm^{-2}$ $s^{-1}$ (minus sign means downward), which is larger than the maximum value of the upward gas diffusion flux $4 \times 10^{12}$ $cm^{-2}$ $s^{-1}$. The net $H_2SO_4$ flux is $-5.7 \times 10^{11}$ $cm^{-2}$ $s^{-1}$. Because of the mass conservation, the net flux is equal to the column-integrated total $H_2SO_4$ production rate we prescribed in the model. The gas diffusion flux of $H_2SO_4$ at the cloud base ($4 \times 10^{12}$ $cm^{-2}$ $s^{-1}$) is seven times larger than the $H_2SO_4$ downward flux from the upper cloud region ($-5.7 \times 10^{11}$ $cm^{-2}$ $s^{-1}$), indicating the strength of the lower cloud cycle is much stronger than the $H_2SO_4$ cycle in its production region in the upper clouds. This result agrees with that in Krasnopolsky & Pollack (1994). Here we highlight several important features in the upper, middle, and lower clouds.

In the upper cloud and haze region (> 56 km), the production of $H_2SO_4$ vapor is balanced by the vapor condensation into the droplets. Because the vapor production rate is large and the condensation is not efficient, the $H_2SO_4$ is highly supersaturated (Figure 5a). Our finding of supersaturated $H_2SO_4$ in the upper clouds is new and different from previous simulations and thus we specifically discussed it in detail in section 3.5. Although the diffusivity is large in this region, the diffusion flux of $H_2SO_4$ vapor is negligible due to the low vapor mixing ratio. As a result, most produced $H_2SO_4$ ends up in the condensed phase (Figure 5c). The condensed $H_2SO_4$ mixing ratio peaks at around 65 km, roughly consistent with the chemical production peak of the $H_2SO_4$ gas (66 km). The former locates slightly lower than the latter due to the particle sedimentation. The upward particle diffusion balances the downward sedimentation near the top domain where condensation is not efficient (Figure 5d).

In the middle cloud region (50 - 56 km), $H_2SO_4$ continues condensing and forming clouds. The supersaturation disappears and the $H_2SO_4$ vapor basically follows the saturation mixing ratio profile (Figure 5a), which agrees with the observations of Pioneer Venus (Kolodner & Steffes, 1998). There is an essential difference between the cloud formation in this region and that in the upper cloud region. In the middle clouds, the condensable $H_2SO_4$ vapor is transported from the lower atmosphere. But in the upper cloud region, the vapor is transported from the chemical source at 66 km. As shown in Figure 5a, the $H_2SO_4$ vapor mixing ratio decreases with increasing altitude below 60 km, indicating an upward diffusion of the $H_2SO_4$ vapor from below. The diffusion convergence of $H_2SO_4$ vapor in this region is balanced by condensation and cloud formation. The condensation rate at 50 km is three times larger than the rate at 66 km, indicating that the droplet size grows rapidly by condensation in the middle clouds.

In the lower cloud region (< 50 km), condensed $H_2SO_4$ droplets accumulate near the cloud base. The droplets falling out of the clouds immediately evaporate, leading to a large $H_2SO_4$ vapor mixing ratio peaking near the cloud base. Eddy diffusion quickly transports the $H_2SO_4$ vapor in both upward and downward directions. As shown in Figure 5, most of the upward transported $H_2SO_4$ vapor is re-condensed into the middle and lower clouds between 50-56 km, which then diffuse and settle down again. In this simulation, we do not include the upward transport of the $H_2SO_4$ below the cloud base by the convection as proposed by Imamura & Hashimoto (2001). Also, because we do not have chemistry in our model, we fixed the lower boundary mixing ratio as 3 ppm to match the positive gradient of the $H_2SO_4$ vapor mixing ratio below the clouds. A coupled cloud-chemical model is needed to simulate the $H_2SO_4$ vapor distribution below the clouds.

### 3.2 $H_2O$

The $H_2O$ cycle in our nominal simulation is represented in Figure 6. The distribution of $H_2O$ vapor mixing ratio is consistent with the observations of ground-based spectroscopy (de Bergh et al., 1991, 1995; Gurwell et al., 2007; Meadows & Crisp, 1996; Sandor & Clancy, 2005). But it is a bit larger than observations of Venus Express (Bertaux et al., 2007; Fedorova et al., 2008) and ground-based spectroscopy (Encrenaz et al., 2013, 2020) above 70 km. In section 4.3, our sensitivity study will show that the $H_2O$ mixing ratio in the mesosphere is sensitive to the $H_2O$ lower boundary value, which might explain the observational variation of the water vapor in Figure 6. As $H_2O$ is transported from the lower atmosphere, the total $H_2O$ mixing ratio decreases with increasing altitude. But because the SVMR of $H_2O$ is significantly larger than that of $H_2SO_4$ (Figure 3), the partitioning of the $H_2O$ in the vapor and condensed phase is different from that of $H_2SO_4$ (Figure 6a). The $H_2O$ vapor mixing ratio follows its SVMR in the clouds and upper hazes. It decreases from the lower atmosphere of 27 ppm to the upper haze region of 4 ppm. The mixing ratio of condensed $H_2O$ has a peak of 3 ppm at 64 km. It reduces to 0.2 ppm at the cloud base. The mixing ratio of $H_2O$ in the condensed phase is about half of the vapor in the $H_2SO_4$ formation region between 64-68 km. Outside this region, the condensed $H_2O$ is significantly smaller than the vapor. This implies that the $H_2O$ in the clouds is mainly transported in the gas phase instead of the condensed phase, opposite to the case of the $H_2SO_4$.

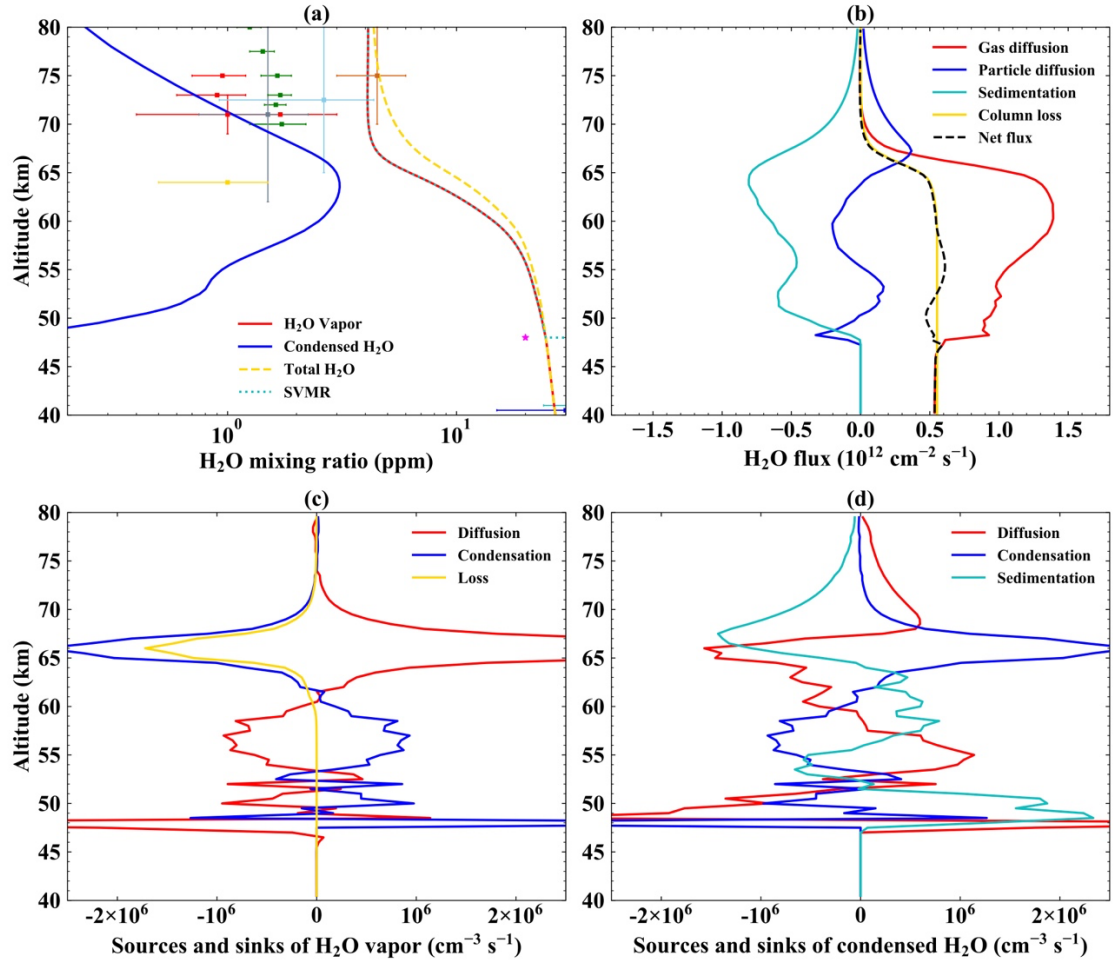

**Figure 6.** The $H_2O$ simulation results in the nominal case. (**a**) the volume mixing ratios of $H_2O$ vapor (red solid line) and the condensed $H_2O$ (blue solid line). The total $H_2O$ mixing ratio is shown as the yellow dashed line. The $H_2O$ saturation mixing ratio is shown as the cyan dotted line (we adopted the SVMR over pure water below the clouds). The observations of $H_2O$ vapor mixing ratio are from Venus Express (red bar, Bertaux et al., 2007; green bar, Fedorova et al., 2008) and ground-based spectroscopy (cyan bar, de Bergh et al., 1991; dark blue bar, de Bergh et al., 1995; grey bar, Encrenaz et al., 2013; yellow bar, Encrenaz et al., 2020; chocolate bar, Gurwell et al., 2007; pink star, Meadows & Crisp, 1996; sky blue bar, Sandor & Clancy, 2005). (**b**) the $H_2O$ fluxes by the gas diffusion (red) and the particle diffusion (dark blue) and sedimentation (cyan). The positive value means an upward flux. The net flux is the black dashed line. The column-integrated loss rate of $H_2O$ (yellow) is calculated from the integration of its chemical loss rate from an altitude level to the top of the atmosphere. (**c**) the source and sink terms of $H_2O$ vapor including diffusion (red), condensation or evaporation (blue), and chemical loss (yellow). (**d**) the source and sink terms of $H_2O$ in condensed phases including

particle diffusion (red), condensation or evaporation (blue), and particle sedimentation (cyan).

The $H_2O$ cycle in our nominal model is very different from the $H_2SO_4$ cycle (Figure 1). $H_2O$ has a large reservoir in the lower atmosphere. As it is diffused upward from the cloud base, it condenses with $H_2SO_4$ in the lower clouds and fall back to the cloud base. But this water cloud cycle between 48-53 km is much weaker than the aforenoted $H_2SO_4$ cloud cycle. Above ~53 km, the upward diffused $H_2O$ vapor is enhanced by about 40% from the evaporation of the droplets precipitating from above. At the $H_2SO_4$ formation (~66 km), about 40% of the lost water vapor reacts with the $SO_3$ to form $H_2SO_4$, and the other is directly condensed into the droplets and then subsequently transported downward by sedimentation. Most of the condensed $H_2O$ is then evaporated before the falling particles reach 53 km as temperature increases. This upper water cycle (53-66 km) is stronger than its lower cycle. Eventually, the cloud droplets fall out of the clouds and evaporate. The fluxes of $H_2O$ (Figure 6b) show that the $H_2O$ is transported upward by gas diffusion and downward by droplet sedimentation in the upper and middle clouds and diffusion in the lower clouds. The upward gas diffusion flux has two peaks of $1.4 \times 10^{12}$ cm$^{-2}$ s$^{-1}$ and $1.0 \times 10^{12}$ cm$^{-2}$ s$^{-1}$ at 66 km and the cloud base, respectively. It is larger than the total downward flux and the net flux is equal to the water column-integrated chemical loss in the formation of $H_2SO_4$. The photolysis of the $H_2O$ is neglected in this study. Thus, the $H_2O$ net flux approaches zero in the upper haze region.

In the upper water cycle, the chemical loss rate of $H_2O$ in our model is fixed the same as the production rate of $H_2SO_4$ (see section 2.3). Both the condensation and the chemical loss of $H_2O$ contribute to the sink of $H_2O$ vapor, leading to a large depletion. The condensation rate peaks at 66 km with a value of $2.7 \times 10^6$ cm$^{-3}$ s$^{-1}$, which is larger than the chemical loss rate in the formation of $H_2SO_4$ by 60%, implying that the molecular ratio between the $H_2O$ and $H_2SO_4$ is about 1.6 in the upper clouds and the resultant cloud acidity is about 80%. The condensed $H_2O$ exhibits a mixing ratio peak at around 64 km where the condensed $H_2SO_4$ also peaks (Figure 5a). Unlike the condensed $H_2SO_4$ that directly settles down all the way to the cloud base, the condensed $H_2O$ gradually evaporates before the falling particles reach 53 km as temperature increases. This is because the response of the $H_2O$ SVMR to the temperature change is different from that of the $H_2SO_4$. Detailed mechanism will be discussed in section 3.4.

Below the upper water condensation-evaporation cycle, the droplets continue to settle down, but the $H_2O$ vapor diffused from below starts to condense into the droplets again. Its condensation rate behaves as a multi-peak structure with a maximum value of $1.3 \times 10^6$ cm$^{-3}$ s$^{-1}$. Below the cloud base, the droplet evaporation releases $H_2O$ vapor, which is also diffused upward and re-condenses. The condensation rate of $H_2O$ in this lower cloud cycle is about an order of magnitude smaller than $H_2SO_4$. The mixing ratio of the condensed $H_2O$ is significantly smaller than the condensed $H_2SO_4$, especially at the cloud base (Figure 6). Thus, the lower water cycle (48-53 km) contributes much less to the lower-cloud mass loading than the $H_2SO_4$ cycle.

The $H_2O$ vapor mixing ratio in our model decreases with altitude below the clouds. The negative gradient of $H_2O$ vapor is consistent with some observations by ground-based spectroscopy (Chamberlain et al., 2013; Meadows & Crisp, 1996) but disagrees with the observations by Venera (11 and 13) (Ignatiev et al., 1997). The positive gradient might imply the possible importance of upward wind transport of $H_2O$ in the equatorial regions as proposed by Imamura & Hashimoto (2001).

3.3 Cloud mass loading and particle size

The simulated vertical structure of the cloud mass loading (Figure 7a) is generally consistent with the observation of Pioneer Venus in the cloud layers (Knollenberg & Hunten, 1980) and the upper hazes below 80 km (Wilquet et al., 2009). Our simulation underestimates the observed mass loading by a factor of two in 52-54 km and overestimates it by a factor of four in 70-73 km. The model also failed to match the large cloud mass peak at around 49 km right above the cloud base. The local clouds are mostly composed of large mode-3 particles at around 49 km (Knollenberg & Hunten, 1980) and small mode-1 particles in the upper hazes (Kawabata et al., 1980; Lane & Opstbaum, 1983; Wilquet et al., 2009, 2012) that our single-mode model did not produce. We will show in section 4.4 that the upper haze in the 70-73 km region can be explained better if we reduce the CCN by a factor of five. The simulated cloud base is at around 48 km. The cloud mass loading generally decreases with increasing altitude except for 56-60 km where the mass loading is roughly constant with height. The droplets accumulate in the lower clouds with a peak mass loading of 15 mg m$^{-3}$ at the cloud base. We did not reproduce the lower haze below 48 km, which might come from a different origin, for instance, the vertical convection in the Venusian troposphere (Schubert et al., 1980) or complex chemical processes in the lower atmosphere (Krasnopolsky, 2007).

Because our model fixed the CCN number density, the cloud mass loading and the particle size are directly linked with each other. The particle diameter profile shows two peaks (Figure 7c) at the cloud base and 66 km with values of 3 μm and 2 μm, respectively. Our particle diameter is consistent with the size of the observed mode 2 particles by Pioneer Venus (Knollenberg & Hunten, 1980) in the lower clouds but slightly smaller than the observation in the upper clouds, though we could not produce the mode 1 and mode 3 particles as the single-mode model does not treat detailed microphysics. We also converted the observed data into "mass-averaged diameter" by distributing the observed cloud mass loading uniformly on each particle (white dashed line in Figure 7c). Our model can reproduce the "mass-averaged diameter" in the lower clouds but underestimate it by 20-30% above 51 km. This disagreement could be caused by our neglect of particle size distribution.

The sedimentation velocity is shown in Figure 7d. It is on the order of $10^{-2}$ cm s$^{-1}$ at the cloud base, agreeing with the estimated sedimentation velocity of the mode 2 particles in McGouldrick & Tsang (2017). Our simulated velocity increases with altitude above 60 km, reaching 0.6 cm s$^{-1}$ at the upper boundary. Efficient particle sedimentation mainly balances the particle diffusion in the upper hazes as the diffusivity also increases with altitude.

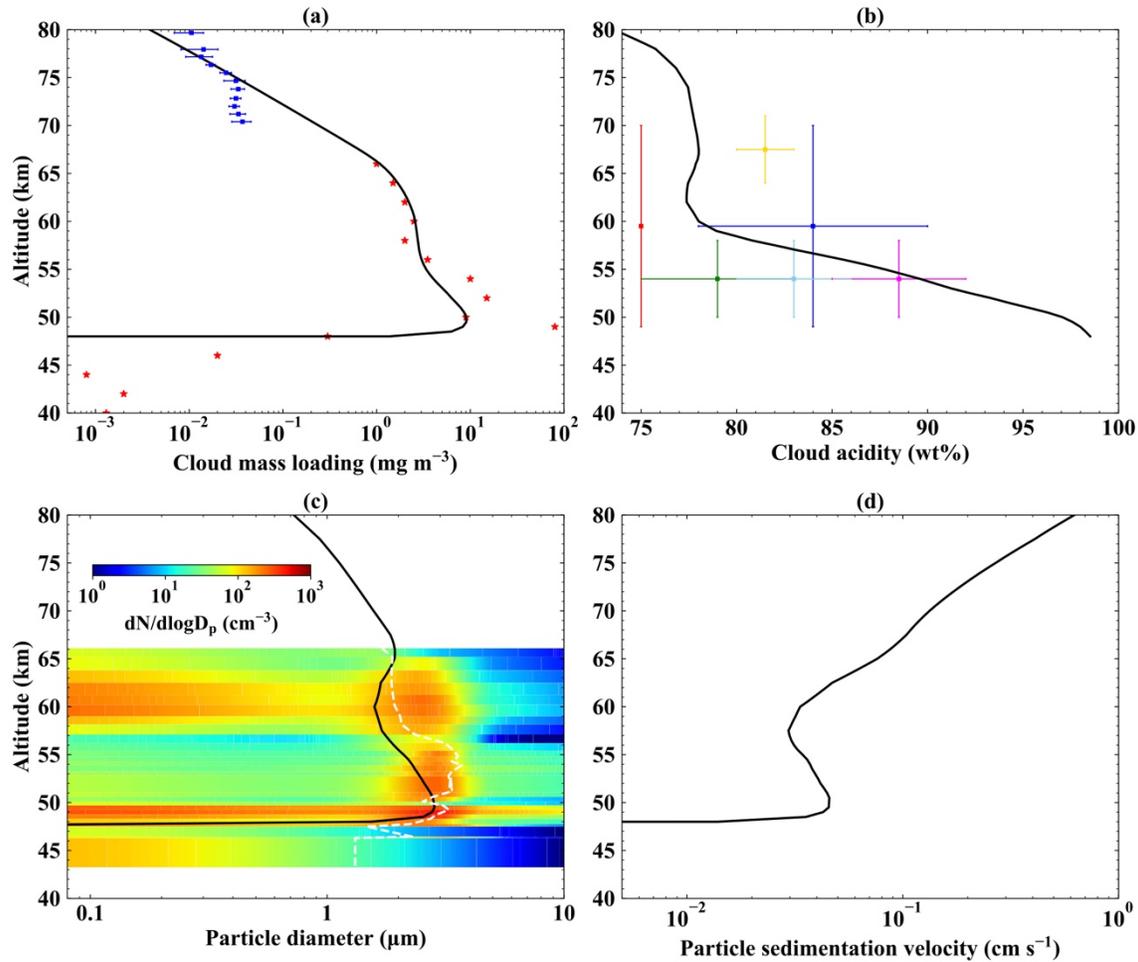

**Figure 7.** The nominal simulation results of clouds: (**a**) The cloud mass loading (black solid line). The observations are from Pioneer Venus (red star, Knollenberg & Hunten, 1980) and Venus Express (dark blue bar, Wilquet et al., 2009); (**b**) The cloud acidity (black solid line) with the estimated values from Venus Express (green, Arney et al., 2014; pink, Barstow et al., 2012; yellow, Cottini et al., 2012; sky blue, McGouldrick et al., 2021) and ground-based spectroscopy (red, Hansen & Hovenier, 1974; dark blue, Pollack et al., 1978). (**c**) The particle diameter (black solid line). The observed particle size distribution with three modes by Pioneer Venus (Knollenberg & Hunten, 1980) is calculated using the fitted analytical functions from Knollenberg & Hunten (1980). The white dashed line represents the "mass-averaged diameter" obtained by distributing the observed total mass uniformly on each particle; (**d**) the particle sedimentation velocity.

3.4 Cloud acidity

Our simulated acidity (Figure 7b) shows a maximum value of 98% at the cloud base and rapidly decreases to the minimum of 78% at 62 km. An inversion exists in 62-70 km, above which altitude the acidity decreases with altitude, forming a peak of 80% near the cloud top. The simulated cloud acidity roughly agrees with the observational estimate ranging from 78% to 92% (Barstow et al., 2012; McGouldrick et al., 2021; Pollack et al., 1978) but is lower than the reported values in Cottini et al. (2012) in the upper clouds and higher than that in Arney et al. (2014) in the middle and lower clouds. Our simulated cloud acidity is also higher than the estimate in Hansen & Hovenier (1974). However, the previously derived cloud acidity values from the observations have large uncertainties. First, none of the above-quoted studies actually retrieved the vertical profiles of cloud acidity—they assumed the acidity is constant throughout the clouds, and the sensed altitudes are highly uncertain. Second, the remote sensing techniques derived the cloud acidity via constraining the imaginary index of the refraction of the droplet, which was measured in limited wavelengths (Arney et al., 2014; Palmer & Williams, 1975). The currently available three data points for 75%, 84.5%, and 95.6% acid (Palmer & Williams, 1975) show strong non-linear dependence of the imaginary index of refraction on the cloud acidity (Barstow et al., 2012) and a simple interpolation would not work for the intermediate acidity. Moreover, because the observed radiance is less sensitive to the cloud acidity than to the cloud opacity, the cloud acidity is generally less constrained (Arney et al., 2014). In addition, since the Venus cloud has large spatial and temporal variations (Titov et al., 2018), the observed cloud acidity might vary as well (McGouldrick et al., 2021).

Cloud acidity has a significant impact on the saturation vapor pressures of both $H_2SO_4$ and $H_2O$. In the upper clouds, the condensed $H_2SO_4$ reduces the SVMR of $H_2O$ and causes the water to condense. The condensation rate of $H_2O$ vapor at 66 km is 60% larger than its chemical loss in the formation of $H_2SO_4$, resulting in cloud acidy of about 80% in the upper clouds. As the particles settle down to the warmer atmosphere below, our model predicts that the condensed $H_2O$, rather than the $H_2SO_4$, would evaporate. Here is the reason. Considering a droplet with 80% acidity that falls from 60 km to 53 km, the $H_2O$ SVMR changes from 10 ppm at 60 km to 100 ppm at 53 km (Figure 3), which greatly exceeds the total $H_2O$ mixing ratio in the Venusian atmosphere (~30 ppm, Bézard et al., 2011; Chamberlain et al., 2013; de Bergh et al., 1991, 1995; Marcq et al., 2006, 2008; Tsang et al., 2010). Therefore, the condensed $H_2O$ inside the droplets has to evaporate back to the vapor phase and increase the droplet acidity. On the other hand, for 80% sulfuric acid solution, the $H_2SO_4$ SVMR changes from $10^{-6}$ ppm at 60 km to $10^{-2}$ ppm at 53 km (Figure 3). This change can be readily compensated by the gas diffusion from below without $H_2SO_4$ evaporation from the droplets. As a result, the droplets become more acidic as they settle down into the hotter and lower atmosphere.

In the lower clouds, because the SVMR of $H_2SO_4$ drops much more quickly with altitude than $H_2O$ (Figure 3), the $H_2SO_4$ vapor is diffused upward from the cloud base condenses more effectively than the water vapor in the lower clouds. Consequently, the lower clouds are even more acidic than the middle clouds. In the lower-cloud region, our simulated acidity is larger than 90% below 55 km and reaches 98% at the cloud base, much larger than the reported values of 79-88% in the lower clouds (Arney et al., 2014; Barstow et al., 2012; McGouldrick et al., 2021). But note that the observationally derived cloud acidity is highly uncertain. From a theoretical point of view in Krasnopolsky (2015), if the lower cloud acidity were 88%, the water vapor should be 600 ppm at 50 km, much larger than the observed $H_2O$ below the clouds (e.g., Arney et al., 2014). As the $H_2O$ SVMR is very sensitive to cloud acidity, future observations of the water vapor in the cloud layer might offer alternative constraints on the cloud acidity.

3.5 The supersaturation of $H_2SO_4$

Our model produces a large supersaturation of $H_2SO_4$ vapor above 60 km (Figure 5a). The vapor mixing ratio is over two orders of magnitude larger than the SVMR from 65 km to 80 km (the model top). This is very different from all previous cloud models on Venus, which basically showed no supersaturation of $H_2SO_4$ at all altitudes except in a narrow region where the $H_2SO_4$ production rate reaches the maximum (Gao et al., 2014; Imamura & Hashimoto, 2001). This difference requires a clear explanation.

The mechanism behind the $H_2SO_4$ supersaturation in our model is straightforward. At 66 km, the chemical production rate of $H_2SO_4$ vapor in our model peaks at about $1.8 \times 10^6$ cm$^{-3}$ s$^{-1}$, which is balanced by condensation as the diffusion is negligible (see Figure 5c). From equations (9) and (10), the condensation rate depends on the particle size and the supersaturation ($q_1 - q_1^{svp}$). Given a typical particle diameter of 1-2 μm in the upper clouds of Venus, $q_1 - q_1^{svp}$ must be on the order of $10^{-6}$ ppm to balance the chemical production rate. Because the upper cloud acidity is about 80%, the saturation vapor mixing ratio of $H_2SO_4$ $q_1^{svp}$ is only on the order of $10^{-8}$ ppm (Figure 3), resulting in a large supersaturation, i.e., $q_1 \approx 10^{-6} \sim 10^{-5}$ ppm or 1-10 ppt (parts per trillion) level. In other words, $H_2SO_4$ is highly supersaturated because the particle condensation is not efficient enough to sequester the chemically produced $H_2SO_4$ vapor back to the SVP in the upper region. Although the $H_2SO_4$ is supersaturated, the 1-10 ppt level vapor mixing ratio is still smaller than the currently observed upper limit of 3 ppb at 85 km from ground-based spectroscopy (Sandor et al., 2012).

Why previous microphysical models (Gao et al., 2014; Imamura & Hashimoto, 2001) did not produce such a broad and large supersaturation in the upper clouds? We found that the vertical profile of the $H_2SO_4$ production rate is the key. In Imamura & Hashimoto (2001), the $H_2SO_4$ production rate profile was calculated from their simple chemical reactions. Figure 8a shows that their profile has a much narrower width in altitude than our profile, and the peak of the profile is also located at 61 km rather than 66 km in our model. Their $H_2SO_4$ production rate basically drops to a very small value above 66 km, more than six orders of magnitude smaller than our adopted rate profile. Because the supersaturation in our model is a result of inefficient condensation to balance the vapor production. With a much smaller chemical production rate above 66 km, $H_2SO_4$ vapor is not supersaturated in the Imamura & Hashimoto model, except at around 61 km where their production rate is still very large (see their Figure 10).

To test this, we used their $H_2SO_4$ production rate profile in our simple condensation model and keep every other parameter the same as our nominal case. The simulated $H_2SO_4$ is not supersaturated above 65 km in this case (Figure 8) and confirms that it is the $H_2SO_4$ production rate controls the $H_2SO_4$ supersaturation. Furthermore, our sensitivity tests of other model parameters such as the eddy diffusivity, CCN number density, and water vapor at the lower boundary (see section 4) consistently show that the $H_2SO_4$ supersaturation is a robust phenomenon if we adopt the $H_2SO_4$ chemical production rate profile from Krasnopolsky (2012). Because the latest photochemical models have been validated against the Venus Express data (Krasnopolsky, 2012; Shao et al., 2020; Zhang, Liang, et al., 2012), we think the $H_2SO_4$ production rate profiles used in our nominal model and sensitivity tests (Figure 4) are more realistic compared with that in Imamura & Hashimoto (2001). We also suggest that future microphysical models of sulfuric acid clouds on Venus should use more robust $H_2SO_4$ production rate profiles from realistic chemical models.

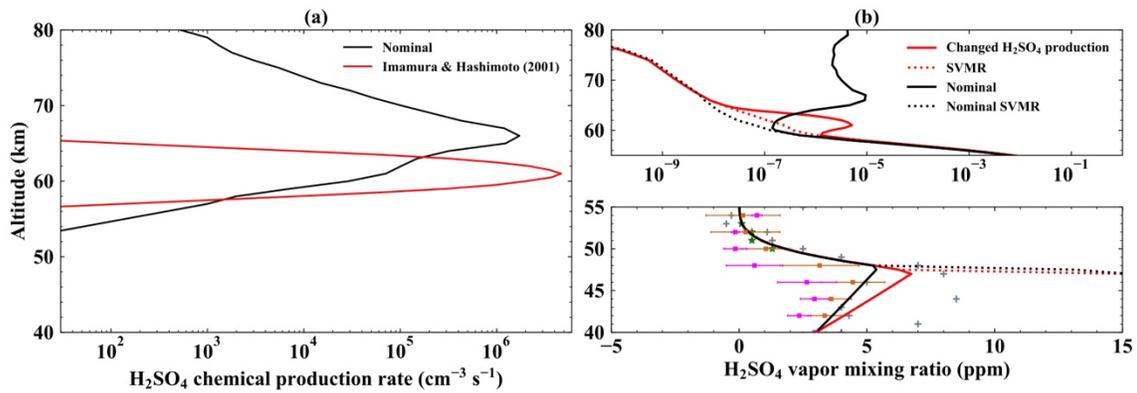

**Figure 8.** (**a**) the $H_2SO_4$ production rate profiles from our nominal simulation (black) and Imamura & Hashimoto (2001, red); (**b**) the simulated $H_2SO_4$ vapor mixing ratios in our nominal simulation (black) and the test (red) where the $H_2SO_4$ production rate profile of Imamura & Hashimoto (2001) is adopted instead. Note that the mixing ratio in the top

right panel is in log scale and the bottom right panel is in linear scale. The observations of the $H_2SO_4$ vapor mixing ratio are the same as in Figure 5a.

## 4 Sensitivity tests

### 4.1 $H_2SO_4$ chemical production rate

The mixing ratio of the $SO_2$ at the cloud top shows a decadal fluctuation observed by Pioneer Venus (Esposito, 1984; Esposito et al., 1988, 1997; Na et al., 1990), Venus Express (Marcq et al., 2012, 2020; Vandaele et al., 2017), and ground-based spectroscopy (Encrenaz et al., 2016, 2019). These observations constrain a cloud-top $SO_2$ mixing ratio range of 1-1000 ppb. Since the $H_2SO_4$ formation depends on the $SO_2$ photochemistry, the fluctuation of the cloud-top $SO_2$ greatly affects the $H_2SO_4$ chemical production rate.

To test the impact of varying $H_2SO_4$ production rates on the clouds, we adopt three $H_2SO_4$ production rate profiles from Shao et al. (2020) by varying the boundary $SO_2$ mixing ratio at 58 km in their model. These $H_2SO_4$ production rate profiles correspond to the cloud-top (70 km) $SO_2$ mixing ratios of 6, 20, 40 ppb, respectively (Figure 4). For reference, in our nominal case the $SO_2$ mixing ratio at the cloud top is about 100 ppb. As the cloud-top $SO_2$ increases, the $SO_2$ photolysis occurs at a higher altitude because the opacity of $SO_2$ increases. As a result, the peak of the production rate is larger and shifts to a higher altitude (Figure 4). We performed a sensitivity study using these three $H_2SO_4$ production rate profiles but keeping other parameters (such as the eddy diffusivity) the same.

The $H_2SO_4$ vapor supersaturation robustly exists in all cases but the supersaturation level changes with the prescribed vapor production rate profile. As we demonstrated in section 3.5, as long as there is a large $H_2SO_4$ production rate in the upper atmosphere, the local cloud/haze condensation cannot sequester the produced $H_2SO_4$, leading to a large supersaturation. We also found that the $H_2SO_4$ production rate has a larger impact on the upper clouds and hazes than the middle and lower clouds (Figure 9). In the upper cloud region, the increase of the $H_2SO_4$ production rate leads to an increasing $H_2SO_4$ vapor mixing ratio. Consequently, the cloud mass loading increases and the cloud top moves higher. Observations have shown that the cloud-top altitude appears higher at lower latitudes (Cottini et al., 2012, 2015; Haus et al., 2013, 2014; Ignatiev et al., 2009; Lee et al., 2012; Zasova et al., 2007). Previous studies attributed this pattern to the distributions of mesospheric temperature field (Lee et al., 2012) or the ultraviolet markings (Ignatiev et al., 2009). Our cloud modeling implies a third possibility. At low latitudes, the cloud-top $SO_2$ is more abundant than the polar region from ground-based observations (Encrenaz et al., 2020). Photochemistry is also more active at low latitudes, resulting in a larger $H_2SO_4$ production rate, a larger cloud mass loading, and a higher cloud top.

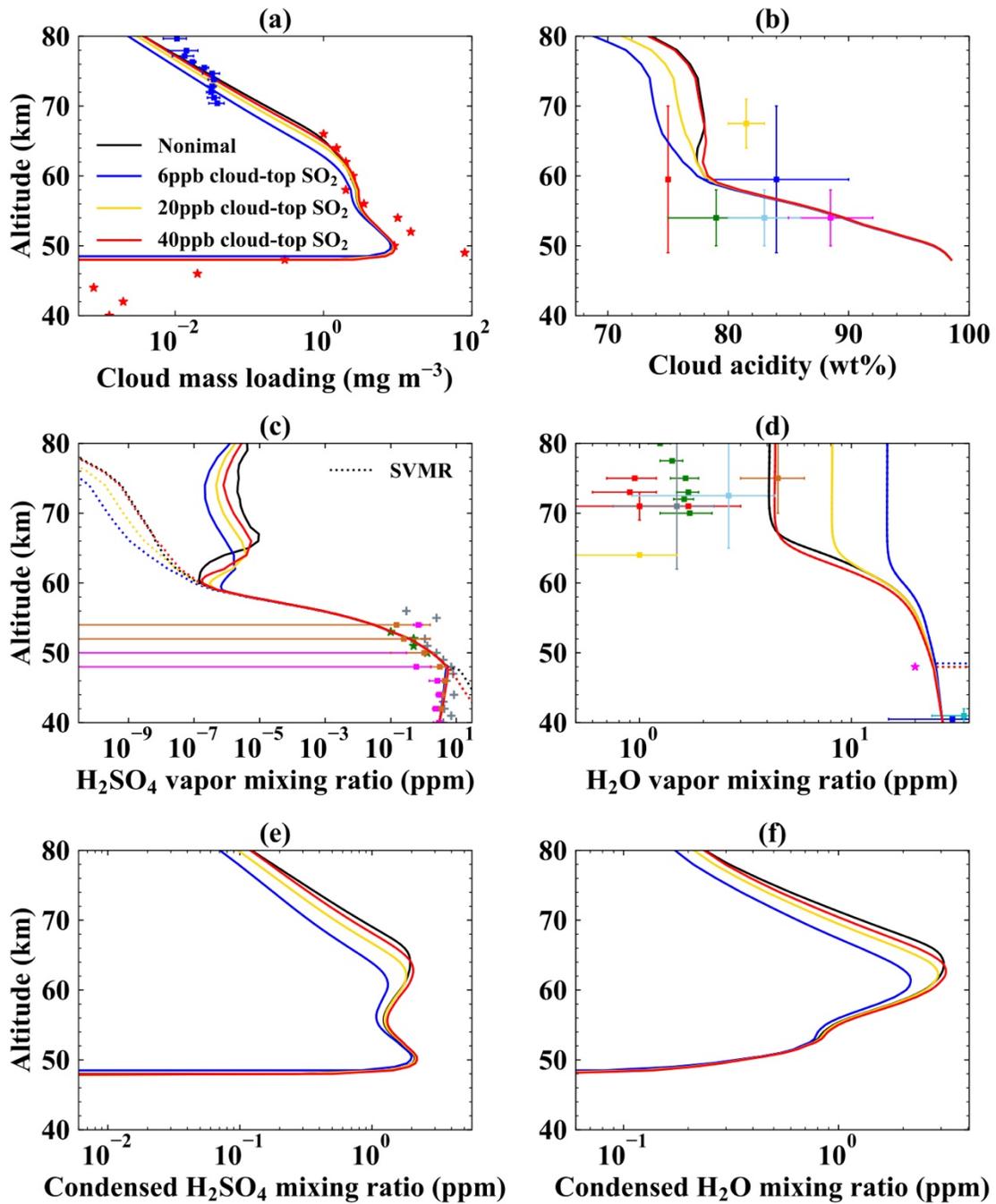

**Figure 9.** The sensitivity tests of the $H_2SO_4$ production rates: (**a**) the cloud mass loading. The observations (bars, stars, and pluses) are the same as in Figure 7a; (**b**) the cloud acidity. The estimations are the same as in Figure 7b; (**c**) $H_2SO_4$ vapor mixing ratio. The dotted lines are the SVMRs. The observations are the same as in Figure 5a; (**d**) $H_2O$

vapor mixing ratio. The observations are the same as in Figure 6a; (**e**) condensed $H_2SO_4$ mixing ratio; (**f**) condensed $H_2O$ mixing ratio.

Our study might also offer another explanation of the anti-correlation between the cloud-top $SO_2$ and $H_2O$ vapor from the ground-based observations in the last decade (Encrenaz et al., 2020). A previous gas chemistry study by Shao et al., (2020) proposed that the gas chemical loss of $H_2O$ could explain the anti-correlation between the cloud-top $SO_2$ and $H_2O$ vapor. Our study suggests that cloud formation might also play an important role. More $SO_2$ in the upper atmosphere would yield a larger $H_2SO_4$ production rate, which, according to Figure 9, would result in a larger cloud acidity in the upper clouds, a lower $H_2O$ SVMR, and a smaller $H_2O$ mixing ratio. But because our cloud model does not couple with gas chemistry, a further detailed study is warranted.
In the middle and lower clouds, the chemistry is negligible (Figure 5c and 6c). Our sensitivity study shows that the change of the sulfuric acid production does not significantly impact the clouds below 56 km. This is expected because the middle and lower clouds are controlled by the strong cloud cycle that is more affected by the dynamics, as we will show below.

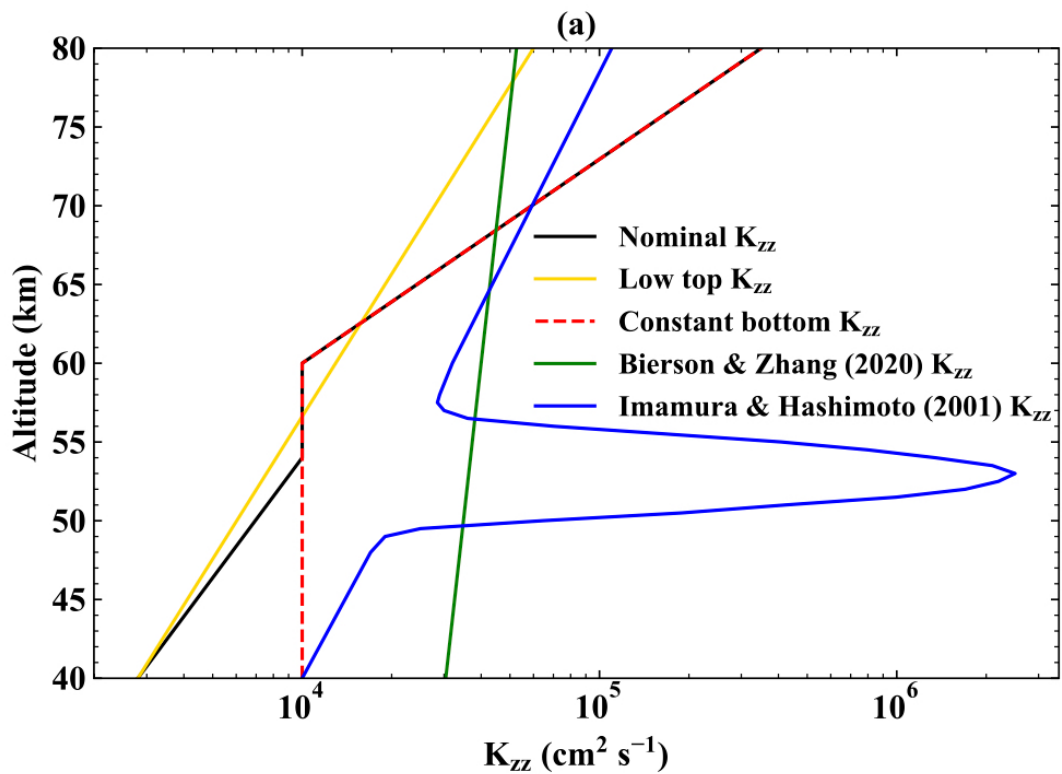

**Figure 10.** The eddy diffusivity profiles tested in this study, including our nominal $K_{zz}$ profile (black), the "low top $K_{zz}$" profile (yellow) that increases exponentially from $2.8×10^3$ cm$^2$ s$^{-1}$ at 40 km to $6×10^4$ cm$^2$ s$^{-1}$ at 80 km, the "constant bottom $K_{zz}$" profile (red) that is the same as the nominal $K_{zz}$ except below 54 km where a constant $K_{zz}$ of $1×10^4$ cm$^2$ s$^{-1}$ is adopted (from Krasnopolsky, 2015), the $K_{zz}$ profile from Bierson & Zhang (2020, green), and that in Imamura & Hashimoto (2001, dark blue). Because the model in Imamura & Hashimoto (2001) only goes up to 70 km, we extrapolated their $K_{zz}$ in the 70-80 km region.

### 4.2 Eddy diffusivity

The eddy diffusivity in the cloud region is currently not well determined. Only a few observational constraints exist (Lane & Opstbaum, 1983; Woo & Ishimaru, 1981). Imamura & Hashimoto (2001) and Gao et al. (2014) adopted an eddy diffusion profile with a strong local maximum of $2×10^6$ cm$^2$ s$^{-1}$ at 53 km to account for unstable stratification observed by the Pioneer Venus (Schubert et al., 1980) and Magellan radio occultations (Hinson & Jenkins, 1995). On the other hand, to explain a large decrease of SO$_2$ in the cloud region, the diffusivity might be low in the stable cloud region where the transport is inhibited (Marcq et al. 2017; Bierson & Zhang 2020). In order to test the impact of eddy diffusion on our simulation results, we performed a sensitivity study using several different $K_{zz}$ profiles (Figure 10) but with a *fixed H$_2$SO$_4$ production rate profile* in our nominal case. We used $K_{zz}$ from Bierson & Zhang (2020), Imamura & Hashimoto (2001), a profile with low $K_{zz}$ values above 60 km ("low top $K_{zz}$"), and a profile with constant Kzz values of $1×10^4$ cm$^2$ s$^{-1}$ below 54 km ("constant bottom $K_{zz}$"). The results are shown in Figure 11.

The $K_{zz}$ sensitivity test offers several important insights into the system. First, the H$_2$SO$_4$ vapor supersaturation ubiquitously exists in all cases, regardless of $K_{zz}$. This is expected because the H$_2$SO$_4$ vapor is controlled by the balance between the production and condensation instead of diffusion. Second, the simulation results in the upper clouds are generally not sensitive to the $K_{zz}$ change above 60 km. This is evidenced by comparing two cases: the nominal case and the "low top $K_{zz}$" case in which the $K_{zz}$ is similar to the nominal profile below 60 km but is much smaller than the other above 60 km. The cloud mass and the water vapor in the low top $K_{zz}$ case are smaller than the nominal case and the cloud acidity is a bit larger. This is because less efficient eddy diffusion would naturally transport less vapor and particles upward.

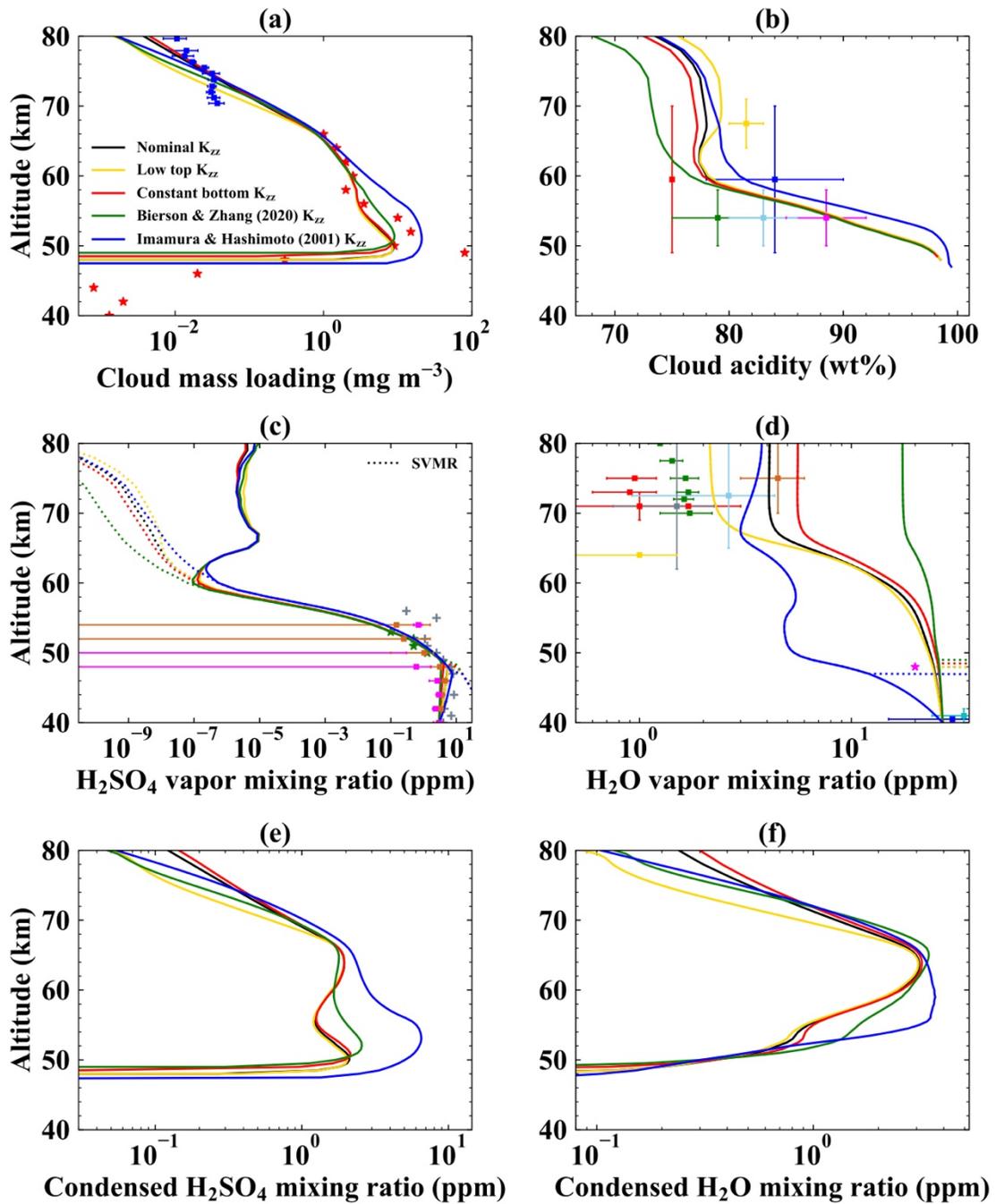

**Figure 11.** Same as Figure 9 but for the sensitivity tests of the eddy diffusivity.

The eddy diffusion in the lower and middle clouds seems to have a profound impact on the cloud system. The $K_{zz}$ in Bierson & Zhang (2020) is the largest below 50 km. In this case, a large amount of water is transported upward and condenses, resulting in a low cloud acidity above 60 km with a large enrichment of water vapor compared with the low top $K_{zz}$ case. A similar behavior is found in the "constant bottom $K_{zz}$" case in which we increase the nominal $K_{zz}$ to $1 \times 10^4$ cm$^2$ s$^{-1}$ below 54 km (Krasnopolsky, 2015). The biggest difference, however, occurs in a case with a $K_{zz}$ peak in the middle cloud layer as that in Imamura & Hashimoto (2001). Such strong diffusivity in the cloud region significantly enhances the $H_2SO_4$ evaporation-condensation cycle (see section 3.1), leading to a large accumulation of condensed $H_2SO_4$ and the large mass flux, which is consistent with Imamura & Hashimoto (2001). Accordingly, the increase of cloud acidity consumes a large amount of water in the lower and middle clouds, leading to a local water depletion at around 53 km where the $K_{zz}$ peak locates (Fig. 11, c, blue line). Because of the enhanced condensation of $H_2SO_4$ and water, our simulation with the $K_{zz}$ from Imamura & Hashimoto (2001) yields much more cloud mass than the other cases. The case shows a slightly larger cloud mass than the observation between 50-60 km (Fig. 11a) but it still cannot match the mode 3 data near the cloud base (about 100 mg m$^{-3}$).

Note that in this test we only changed the eddy diffusivity while keeping other input parameters the same. In reality when the eddy diffusivity increases, the cloud-top $SO_2$ mixing ratio is also likely to increase (Marcq et al., 2012, Bierson and Zhang 2020). As a result, the $H_2SO_4$ production rate could increase and the water vapor mixing ratio might decrease (see section 4.1). A model with coupled chemistry and cloud formation is needed to further explore the details.

4.3 $H_2O$ vapor mixing ratio in the lower atmosphere

The observed $H_2O$ vapor is 30±10 ppm at 30-45 km by ground-based spectroscopy (Arney et al., 2014; Chamberlain et al., 2013; de Bergh et al., 1991, 1995; Marcq et al., 2006) and Venus Express (Barstow et al., 2012; Bézard et al., 2011; Marcq et al., 2008; McGouldrick et al., 2021; Tsang et al., 2010). Here we test different $H_2O$ vapor mixing ratios of 23 ppm, 27 ppm (nominal), and 34 ppm at the lower boundary, respectively, and their impact on the simulation results (Figure 12).

We find that decreasing the lower boundary $H_2O$ vapor greatly decreases the $H_2O$ vapor in the mesosphere above 60 km as less water is transported upward and condenses. When we set the lower boundary value as 23 ppm, the $H_2O$ vapor mixing ratio is less than 1 ppm above 70 km, slightly smaller than the ppm level observations from Venus Express (Bertaux et al., 2007; Fedorova et al., 2008) and ground-based spectroscopy (Encrenaz et al., 2013, 2020).

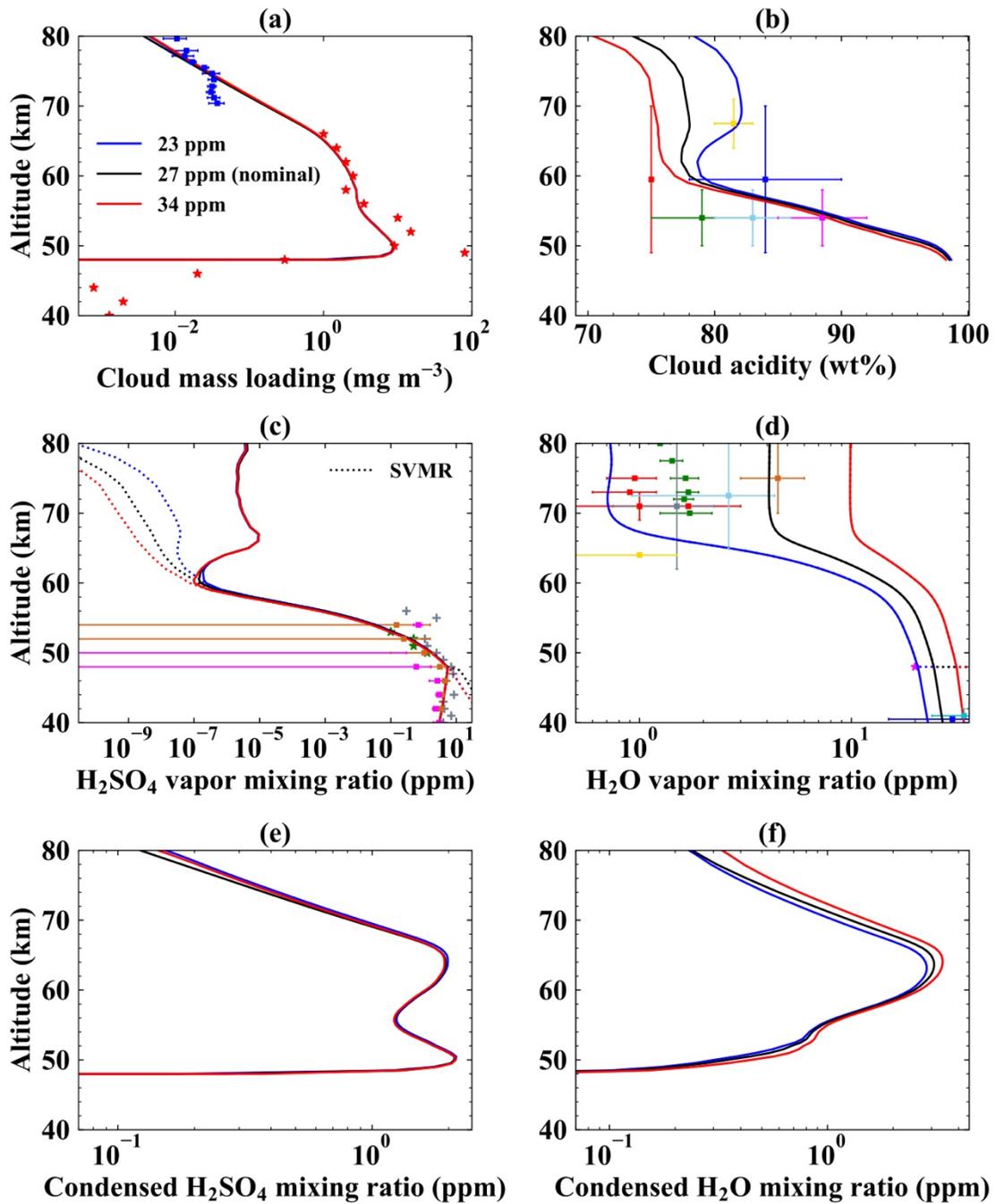

**Figure 12.** Same as Figure 9 but for the sensitivity tests of the $H_2O$ vapor mixing ratio in the lower atmosphere.

Increasing the lower boundary $H_2O$ vapor decreases the cloud acidity as more water condenses with respect to $H_2SO_4$ in the droplets. Our simulations show that increasing the lower boundary $H_2O$ vapor does not significantly affect the cloud mass loading. This seems to contradict the observations by Tsang et al. (2010) that found positive correlations between the sub-cloud $H_2O$ vapor mixing ratio and the lower-cloud opacity by observations of Venus Express. The reason is that in our test we only varied the $H_2O$ vapor mixing ratio while keeping the eddy diffusion unchanged. In reality, the water vapor mixing ratio might vary with the atmospheric dynamics such as convection as proposed by Tsang et al. (2010). Also, Kopparla et al. (2020) proposed that the sub-cloud $H_2O$ vapor mixing ratio has strong influences on the heating of the cloud base and the diffusivity in the clouds. If gas diffusion changes with the water vapor, our simulated cloud mass loading could change too (see section 4.2). A dynamical model with cloud formation and radiative feedback is required for future investigation.

### 4.4 CCN number density

We also performed a sensitivity study to investigate the impact of CCN number density on the simulation results. We vary the CCN number density profile in our nominal case (Figure 2) by a factor of five to encompass the variation range of the droplet number density from LCPS observations (Knollenberg & Hunten, 1980) and the upper haze estimations (Wilquet et al., 2009). Our simulations show that increasing (decreasing) the CCN number density by a factor of five leads to a decrease (an increase) of cloud mass loading by a factor of two at all altitudes. The SVMRs and cloud acidity are almost not affected (Figure 13) because changing the CCN number density ($n_p$) also change the condensation rates of both $H_2SO_4$ and $H_2O$ in proportion to each other (equations 9 and 10). In the region where the vapors follow the SVMRs, the vapor mixing ratios do not change.

In the $H_2SO_4$ supersaturation region above 60 km, the condensation rate should balance our prescribed production rate and should not change in these cases. According to equation (9), the condensation rate is proportional to the product of the CCN number density and the supersaturated $H_2SO_4$ vapor pressure—which is roughly equal to the $H_2SO_4$ vapor pressure itself as the SVMR of $H_2SO_4$ is negligible in that region. As a result, the $H_2SO_4$ vapor mixing ratio is inversely proportional to the CCN number density. As our simulations show, if CCN number density increases by a factor of five, the $H_2SO_4$ vapor mixing ratio decrease by a factor of three because the reduced particle size further decreases the condensation efficiency—but the $H_2SO_4$ vapor is still highly supersaturated (Figure 13).

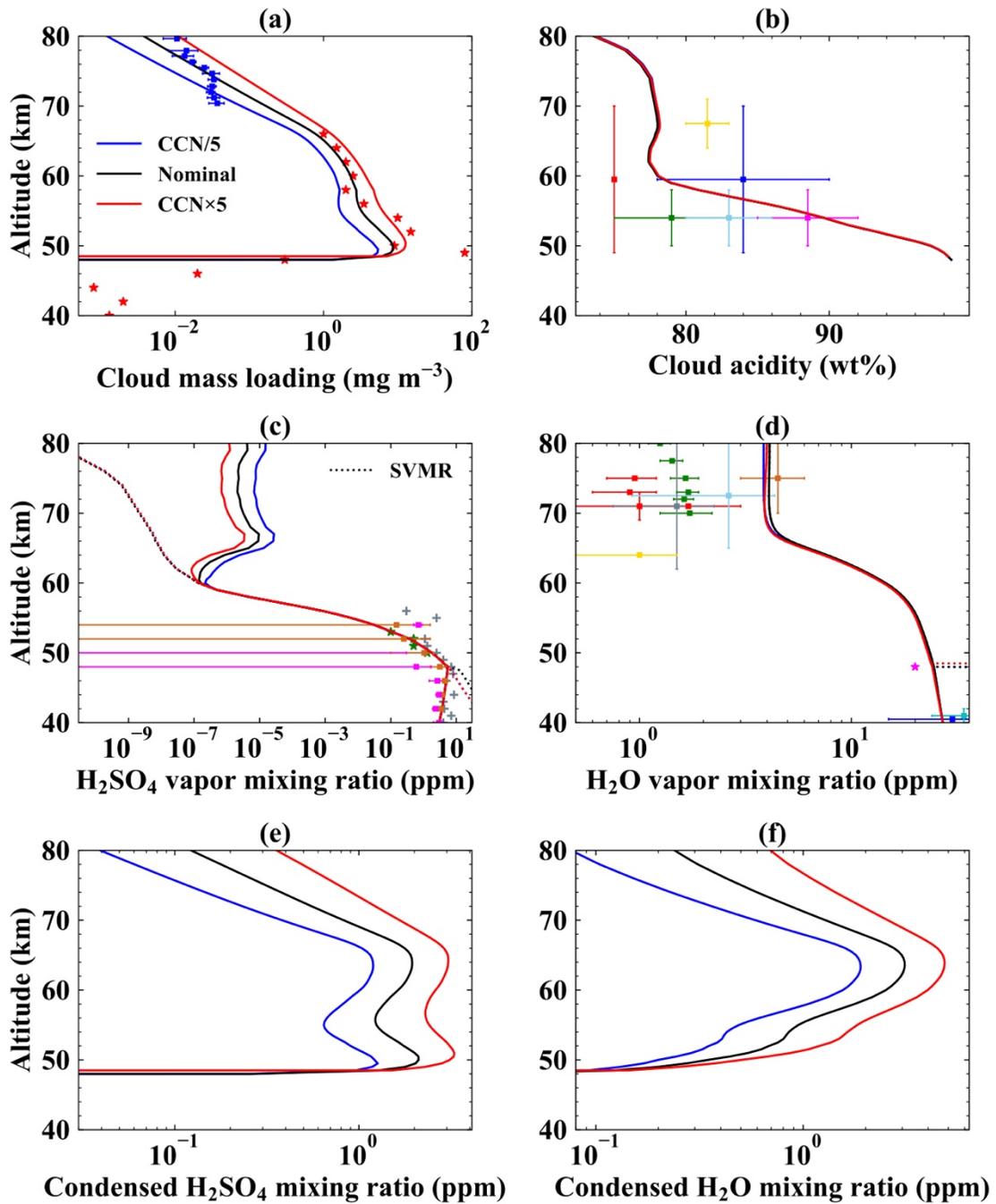

**Figure 13.** Same as Figure 9 but for the sensitivity tests of the CCN number density.

## 5 Comparison to previous model results

Here we briefly compare our simple model with previous models of Venus clouds. The biggest difference between our model and previous models is the $H_2SO_4$ supersaturation above 60 km that we have discussed in detail in section 3.5. In this section, we mainly focus on other differences.

The model by Krasnopolsky & Pollack (1994)—and updated by Krasnopolsky (2015)—solved the coupled problem of diffusion and condensation for the $H_2SO_4$-$H_2O$ system in Venus clouds. Their governing equation was motivated by the conservation of the hydrogen flux in the atmosphere, establishing a useful framework to understand cloud acidity and vapor distribution. Their simulated cloud acidity distribution agreed with our results. Both our and their models predict larger acidity than the observations in the lower clouds. As noted, the retrieved acidity might be subjected to large uncertainties. If the observations are correct, the data might indicate some contamination in the cloud droplets by hydroxide salts (Rimmer et al., 2021). On the other hand, Krasnopolsky & Pollack (1994) and Krasnopolsky (2015) assumed LTE between the droplets and vapors, which meant the vapor partial pressures were always equal to their SVPs. Thus, their model could not produce $H_2SO_4$ supersaturation in the upper atmosphere.

The models by James et al. (1997), Imamura & Hashimoto (2001), and Gao et al. (2014) adopted a one-dimensional microphysical and vertical transport framework to simulate the distributions of the particle number densities and size distributions. Their models are more complicated than our simple framework with fixed particle number density and a single particle size mode. James et al. (1997) and Imamura & Hashimoto (2001) only focused on the middle and lower clouds and solved both $H_2SO_4$ and $H_2O$ vapor and their interaction with the cloud acidity, while Gao et al. (2014) focused on the upper haze layer and adopted a fixed $H_2O$ profile and the acidity. Even our treatment of cloud physics is very different from the microphysical models, our model results are generally consistent with their results in the middle and lower clouds and all models can explain the observed vapor, cloud mass loading, and cloud acidity.

Aside from the $H_2SO_4$ supersaturation above 60 km, the model differences come from various reasons. The microphysical models adopted a very large eddy diffusivity ($1\times10^6$ cm$^2$ s$^{-1}$) in the lower clouds and we have shown in section 4.2 the eddy diffusivity influences the cloud mass and the water vapor profile. Thus, our simulated sedimentation and diffusion fluxes are rather different from theirs. For example, below 58 km—the top boundary of the model by James et al. (1997), the shape of their condensed and vapor $H_2SO_4$ fluxes agrees with that in our model. But the cloud downward sedimentation flux in their nominal model is larger than $10^{-8}$ g cm$^{-2}$ s$^{-1}$ at the cloud base, while our model yields $10^{-9}$ g cm$^{-2}$ s$^{-1}$, consistent with the model results in Krasnopolsky & Pollack (1994). Both the models in Imamura & Hashimoto (2001) and Gao et al. (2014) emphasized the importance of upward wind transport to explain the particle size distributions but we did not have these physics in our model, nor did in James et al. (1997). Moreover, our simulated negative $H_2O$ vapor gradient below the clouds is consistent with ground-based observations (Chamberlain et al., 2013; Meadows & Crisp, 1996). But it disagrees with the observations of Venera (11 and 13) (Ignatiev et al., 1997), which might indicate the possible importance of upward winds in the equatorial region. Because our model setup is so different, we do not tend to make a very detailed comparison here.

Another microphysical model was built by McGouldrick (2017), which argued that there was no solid evidence for the existence of involatile particles in 60-80 km. They proposed that the droplets might be formed via binary homogeneous nucleation, which requires sufficiently high vapor supersaturation. Our framework is different because we assumed pre-existing CCN that could either form via homogeneous or heterogeneous nucleation processes. Our simulations found that even with the abundant CCN, there is still strong $H_2SO_4$ vapor supersaturation above 60 km, allowing the homogeneous nucleation to proceed and supporting the idea in McGouldrick (2017). The McGouldrick (2017) model underestimated the cloud mass loading and overestimated the particle size in the upper clouds, which were attributed to the overestimation of the $H_2SO_4$ chemical production that led to the excessive supersaturation, the rapid growth of the particles, and the fast sedimentation. In our model, increasing the $H_2SO_4$ production rate in the upper clouds results in increasing both cloud mass loading and particle size. This is because we prescribed the CCN profile in the model. While in their model, the CCN can be efficiently formed through homogeneous nucleation under high supersaturation. As a result, the particles rapidly grow and settle down.

Our simulated cloud acidity distribution is different from that in Rimmer et al. (2021), which proposed that the $SO_2$ inside the clouds is depleted through dissolving hydroxide salts into the droplets. Their model couples the gas chemistry and simple cloud formation with aqueous chemistry. The inclusion of hydroxide salts causes the decreasing cloud $pH$ (hydrogen ion concentration) with increasing altitude in their model, contradicting our simulated cloud acidity distribution (Figure 7). In the hot lower cloud layers, $H_2O$ vapor is difficult to condense unless its SVMR is reduced by the strong hydrophilicity of the clouds or aerosols. It should be pointed out that, because their clouds include a large number of hydroxide salts especially in the lower cloud region, the impact on the $H_2O$ SVMR is largely uncertain due to a lack of experimental data. In addition, their $H_2SO_4$ vapor mixing ratio increases with altitude in 20-80 km (see Figure 7 in their paper). This seems inconsistent with the observations of Magellan Mission (Kolodner & Steffes, 1998), Akatsuki Mission (Imamura et al., 2017), and Venus Express (Oschlisniok et al., 2012, 2021) where the observed $H_2SO_4$ vapor mixing ratio decreases with increasing altitude in the cloud layers.

Besides, the $H_2O$ activity in the cloud layer—the ratio of the $H_2O$ vapor mixing ratio to the SVMR over pure condensed water—is important for the habitability in the Venusian atmosphere (Hallsworth et al., 2021). Our calculated $H_2O$ activity profile exponentially increases from 0.000025 at 40 km to the peak of 0.003 at 60 km, then decreases to its minimum of 0.0003 at 74 km. The distribution generally agrees with that in Hallsworth et al. (2021). Since the maximum $H_2O$ activity of 0.003 is far below the 0.585 limit for known extremophiles (Stevenson et al., 2016), the Venus clouds seem uninhabitable.

**6 Conclusions**

In this work, we have developed a 1D cloud model to solve the coupling system of eddy diffusion, chemical production and loss, condensation, and droplet sedimentation for species $H_2SO_4$ and $H_2O$ in the clouds and hazes in the atmosphere of Venus. Our 1D model only considers the condensation mode rather than other detailed microphysics such as nucleation process. Yet the model is able to simultaneously explain four critical observations. The $H_2SO_4$ vapor mixing ratio agrees with the observations of Magellan Mission (Kolodner & Steffes, 1998), Akatsuki Mission (Imamura et al., 2017), Venus Express (Oschlisniok et al., 2012, 2021) in the middle and lower clouds. The $H_2O$ vapor mixing ratio agrees with the observations of ground-based spectroscopy (de Bergh et al., 1991, 1995; Gurwell et al., 2007; Meadows & Crisp, 1996; Sandor & Clancy, 2005) but is a bit higher than observations of Venus Express (Bertaux et al., 2007; Fedorova et al., 2008) and ground-based spectroscopy (Encrenaz et al., 2013, 2020) above 70 km. Note that it is sensitive to the water vapor at the lower boundary of our model. The cloud acidity generally agrees with previous crude estimates from observations ranging from 78% to 92% (Barstow et al., 2012; McGouldrick et al., 2021; Pollack et al., 1978) but is lower than that of Cottini et al. (2012) in the upper clouds, higher than that in Arney et al. (2014) and Hansen & Hovenier (1974). But note that the reported cloud acidity values from observations have large uncertainties. The mass loading is generally consistent with the in-situ observation of Pioneer Venus in the cloud layers (Knollenberg & Hunten, 1980) and the estimated value in the upper hazes above 74 km (Wilquet et al., 2009). But our model overestimates the mass loading by a factor of four in 71-74 km, underestimates the observed mass loading by a factor of 2 at around 52-54 km, and the simulated cloud mass is one order of magnitude smaller than the observed mode-3 particles at around 49 km. Our simulated particle diameter generally agrees with the mode-2 particle size and the "mass-averaged diameter" based on the Pioneer Venus data (Knollenberg & Hunten, 1980) in the lower clouds, except that the simulated size is 20-30% smaller in the middle clouds. The particle sedimentation velocity simulated in our model increases with altitude from $10^{-2}$ cm s$^{-1}$ in the cloud base to $10^{-1}$ cm s$^{-1}$ in the upper hazes.

Our simple model results generally agree with previous detailed microphysical model results (Gao et al., 2014; Imamura & Hashimoto, 2001; James et al., 1997) although our model runs much faster than detailed microphysical models. One pronounced difference between our model results and previous work is that the $H_2SO_4$ vapor above 60 km is largely supersaturated by over two orders of magnitude in our model while it follows the saturation vapor curves in previous models. We found that the key difference is the $H_2SO_4$ vapor production via chemistry in the upper atmosphere. In the latest photochemical models on Venus such as in Krasnopolsky (2012), the $H_2SO_4$ production rate above 65 km is much larger than the one adopted in previous microphysical cloud models. Consequently, the cloud condensation is not efficient enough to sequester the produced $H_2SO_4$ vapor, resulting in a large supersaturation. Our predicted $H_2SO_4$ vapor mixing ratio in the mesosphere is 1-10 ppt, which is still smaller than the currently observed upper limit of 3 ppb at 85 km from ground-based spectroscopy (Sandor et al., 2012). Future observations might push the to a lower detection limit to test our model prediction. Our study suggests that future cloud modeling studies on Venus should pay attention to the vertical profile of the $H_2SO_4$ vapor production rate in the mesosphere from gas chemistry. On the other hand, some previous gas chemical models of Venus assumed the $H_2SO_4$ vapor under LTE conditions (e.g., Zhang et al. 2010, 2012), which might also need to be revised.

Our simple model allows an easy diagnosis of the mechanisms behind the cloud cycles. The main cycles have been illustrated in Figure 1. The 53-56 km region seems to separate the vapor-cloud system in our model into two distinguished regions. The upward diffused $H_2SO_4$ vapor is re-condensed below this layer and the settling condensed $H_2O$ is mainly evaporated above this layer. This might lead to the distinct cloud mass loading difference between the upper and middle clouds as observed by Pioneer Venus (Knollenberg & Hunten, 1980). The chemically produced $H_2SO_4$ vapor is rapidly condensed into the droplets. The $H_2SO_4$ in the cloud regions is mainly transported downward in the condensed phase and transported upward in vapor. The condensed $H_2SO_4$ is rapidly evaporated once the particles fall below the cloud base. Part of $H_2SO_4$ vapor is transported upward by diffusion and condensed back into the droplets. The upward flux of $H_2SO_4$ vapor at the cloud base is seven times larger than the downward flux of condensed $H_2SO_4$ at 66 km that is controlled by the chemical production rate. The downward transported condensed $H_2SO_4$ and the upward transported $H_2SO_4$ vapor near the cloud base lead to the accumulation of the particles in the lower clouds, resulting in increasing cloud mass loading with decreasing altitude with a peak of 12 mg m$^{-3}$ at the cloud base. The rest of the $H_2SO_4$ vapor is transported downward, which will be thermally decomposed in the lower atmosphere as observed by Magellan (Jenkins et al., 1994) and simulated by chemical models (Bierson & Zhang, 2020; Krasnopolsky, 2007).

The $H_2O$ cycle is different. The water vapor is transported upward from the lower atmosphere by diffusion. It is condensed into the droplets in the upper clouds, leading to a cloud acidity peak of 80% at 70 km. The condensed $H_2O$ mixing ratio is significantly smaller than condensed $H_2SO_4$ in the lower clouds due to the larger $H_2O$ SVMR. Most of the condensed $H_2O$ in the upper clouds is evaporated before particles fall to the middle clouds as the temperature increases. Below this upper water cycle (53-66 km), re-condensation of the evaporated $H_2O$ from the cloud base is weak. Therefore, the condensed $H_2O$ mixing ratio in the lower clouds is small, contributing much less to the lower-cloud mass loading than $H_2SO_4$. Also, the SVMR of $H_2SO_4$ decreases much more dramatically with altitude than $H_2O$ above the cloud base, more $H_2SO_4$ condenses in the lower clouds. As a result, the cloud acidity rapidly decreases from the cloud base value of 98% to 79% at 62 km.

Our simple model also allows an efficient exploration of the parameter space for input parameters. We performed sensitivity tests on the $H_2SO_4$ chemical production rate, the eddy diffusivity, the sub-cloud $H_2O$ vapor mixing ratio, and the CCN number density. We list the dependence of each observable on the input parameters in Table 2. Increasing the $H_2SO_4$ production rate enhances the $H_2SO_4$ supersaturation and cloud mass loading in the upper clouds and hazes but has less impact on the middle and lower clouds. A larger $H_2SO_4$ production rate also increases the cloud acidity and reduces the $H_2O$ mixing ratio. The $H_2SO_4$ production rate might also play a role in the latitudinal variations of the cloud-top altitude and the anti-correlation between the $SO_2$ and $H_2O$ vapor in the upper clouds. In general, a larger eddy diffusivity in the upper hazes leads to a larger $H_2O$ mixing ratio, smaller droplet acidity, and larger cloud mass as more $H_2O$ is transported upward from the lower atmosphere. But the vertical profile of the eddy diffusivity is also important. For example, if there is a local maximum of eddy diffusivity in the middle cloud layer as assumed in previous microphysical models (e.g., Imamura & Hashimoto, 2001), the $H_2SO_4$ cloud cycle is greatly enhanced. As a result, both cloud acidity and lower-cloud mass loading significantly increase, and the water vapor is sequestered in the cloud region. A larger sub-cloud $H_2O$ vapor mixing ratio naturally leads to a larger $H_2O$ vapor mixing ratio and lower cloud acidity in the entire atmosphere, but it does not impact the cloud mass or the $H_2SO_4$ vapor distribution. Increasing the CCN number density would increase the condensation rate and thus increase the cloud mass and reduce the $H_2SO_4$ vapor mixing ratio in the mesosphere, while the cloud acidity and $H_2O$ vapor are not influenced. All our sensitivity cases robustly predict large supersaturation of the $H_2SO_4$ vapor above 60 km.

**Table 2.** *Model dependence on input parameters.*

| Input | Cloud mass | Acidity | $H_2SO_4$ vapor | $H_2O$ vapor |
|---|---|---|---|---|
| $H_2SO_4$ production rate | + | + | + | - |
| Eddy diffusivity[a] | + | - | × | + |
| $H_2O$ at the lower boundary | × | - | × | + |
| CCN number density | + | × | - (above 60 km) | × |

*Note.* "+": positive correlation. "-": negative correlation. "×": not sensitive.
[a]Eddy diffusion is more complicated as the vertical profile of eddy diffusivity matters. See section 4.3 for more details.

Future observations by ground-based telescopes and recently selected missions (DAVINCI+, VERITAS, and EnVision) could put more constraints on the $H_2SO_4$ vapor mixing ratio in the mesosphere and the $H_2O$ vapor mixing ratio in the middle and lower clouds, which are essential for testing existing cloud models and constraining the cloud processes on Venus. Laboratory studies are needed to provide a refractive index of concentrated sulfuric acid to help retrieve better cloud acidity from the near-infrared spectra. Future modeling efforts should be put forward to understand the interactions between the chemistry, dynamics (tracer transport), and cloud formation in the $H_2SO_4$-$H_2O$ gas-cloud system on Venus.

**Appendix A: Numeral Scheme**

Here we describe our numerical scheme to solve the continuity equations for $H_2SO_4$ and $H_2O$ in the main text (equation 1):

$$\frac{\partial n}{\partial t} = S_{diff} + S_{chem} + S_{cond} + S_{sed}. \tag{A1}$$

Following the derivations in section 2, we obtained the final equation for gas:

$$n_{atm} \frac{\partial q_i^g}{\partial t} = -\frac{\partial}{\partial z}\left(-K_{zz} n_{atm} \frac{\partial q_i^g}{\partial z}\right) + C_i \left(q_i^g - q_i^{svp}\right) + P_i - L_i \tag{A2}$$

And the final equation for condensates:

$$n_{atm} \frac{\partial q_i^c}{\partial t} = -\frac{\partial}{\partial z}\left(-K_{zz} n_{atm} \frac{\partial}{\partial z}\left(\frac{M_p}{n_{atm}}\right)\right)\left(\frac{M_i q_i^c}{\sum_i (M_i q_i^c)}\right)\frac{N_a}{M_i} - C_i \left(q_i^g - q_i^{svp}\right) - \frac{\partial}{\partial z}(v q_i^c n_{atm}) \tag{A3}$$

where the superscript $g$ and $c$ denote gas and condensed species, respectively, the subscript $i = 1, 2$ represents to $H_2SO_4$ and $H_2O$, respectively, $t$ and $z$ represent the time and the vertical altitude, respectively, $S$ is the source and sink term, $n$ is the number density of the species, $q$ is the volume mixing ratio, $n_{atm}$ is the atmospheric density, $K_{zz}$ is the eddy diffusion coefficient, $P$ and $L$ are chemical production and loss rate, respectively, $M_p$ is the cloud mass loading (equation 16 in the main text), $M$ is the molar mass, $N_a$ is the Avogadro constant, $q^{svp}$ is the saturation vapor mixing ratio (Figure 3), $v$ is the particle sedimentation velocity (equation 17 in the main text), $C_i$ is the condensation rate coefficient. We have

$$C_1 = \frac{-2\pi n_p D_1 M_1 n_{atm} D_p}{(M_1 + M_2 m)\left(1 + \frac{2\lambda_1}{D_p \alpha}\right)} \tag{A4}$$

for $H_2SO_4$ and

$$C_2 = \frac{-2\pi n_p D_2 M_2 n_{atm} D_p}{\left(\frac{M_1}{m} + M_2\right)\left(1 + \frac{2\lambda_2}{D_p \alpha}\right)} \tag{A5}$$

for $H_2O$, where $D$ is the molecular diffusion coefficient (equation 3 in the main text), $n_p$ is the cloud condensation nuclei (CCN) number density, $D_p$ is the particle diameter (equation 6 in the main text), $m = q_2^c/q_1^c$ is the molecular ratio of two condensed species, $\lambda$ is the mean free path (equation 4 in the main text), $\alpha$ is the accommodation coefficient (Seinfeld & Pandis, 2016, p.500-502).

1. Solving for gases.

For the gas species, the continuity equation (A2) at the *jth* layer (*j=1…N*) and *kth* timestep can be discretized in an implicit form:

$$n_{atm(j)} \frac{q^g_{i(j,k+1)} - q^g_{i(j,k)}}{\Delta t}$$
$$= \frac{K_{zz(j+1)}n_{atm(j+1)} - K_{zz(j-1)}n_{atm(j-1)}}{2\Delta z} \cdot \frac{q^g_{i(j+1,k+1)} - q^g_{i(j-1,k+1)}}{2\Delta z}$$
$$+ K_{zz(j)}n_{atm(j)} \frac{q^g_{i(j+1,k+1)} - 2q^g_{i(j,k+1)} + q^g_{i(j-1,k+1)}}{\Delta z^2}$$
$$+ C_{i(j,k)}\left(q^g_{i(j,k+1)} - q^{SVP}_{i(j,k)}\right) + P_{i(j)} - L_{i(j)}$$

$$(A6)$$

where $\Delta t = t_{k+1} - t_k$, $\Delta z = 50$ m. This equation is reformed as:

$$-\frac{n_{atm(j)}q^g_{i(j,k)}}{\Delta t} + C_{i(j,k)}q^{SVP}_{i(j,k)} - P_{i(j)} + L_{i(j)}$$
$$= \frac{1}{4\Delta z^2}\left(K_{zz(j+1)}n_{atm(j+1)} - K_{zz(j-1)}n_{atm(j-1)} + 4k_{zz(j)}n_{atm(j)}\right)$$
$$\cdot q^g_{i(j+1,k+1)} + \left(-\frac{2K_{zz(j)}n_{atm(j)}}{\Delta z^2} + C_{i(j,k)} - \frac{n_{atm(j)}}{\Delta t}\right)\cdot q^g_{i(j,k+1)}$$
$$+ \frac{1}{4\Delta z^2}\left(-K_{zz(j+1)}n_{atm(j+1)} + K_{zz(j-1)}n_{atm(j-1)} + 4K_{zz(j)}n_{atm(j)}\right)$$
$$\cdot q^g_{i(j-1,k+1)}.$$

$$(A7)$$

In $t = t_{k+1}$ and $z_1 < z < z_N$, we let

$$a_j = \frac{1}{4\Delta z^2}\left(-K_{zz(j+1)}n_{atm(j+1)} + K_{zz(j-1)}n_{atm(j-1)} + 4K_{zz(j)}n_{atm(j)}\right), \quad (A8)$$

$$b_j = \left(-\frac{2K_{zz(j)}n_{atm(j)}}{\Delta z^2} + C_{i(j,k)} - \frac{n_{atm(j)}}{\Delta t}\right), \quad (A9)$$

$$c_j = \frac{1}{4\Delta z^2}\left(K_{zz(j+1)}n_{atm(j+1)} - K_{zz(j-1)}n_{atm(j-1)} + 4k_{zz(j)}n_{atm(j)}\right), \quad (A10)$$

$$w_j = -\frac{n_{atm(j)}q^g_{i(j,k)}}{\Delta t} + C_{i(j,k)}q^{SVP}_{i(j,k)} - P_{i(j)} + L_{i(j)}. \quad (A11)$$

The system can be written as a tridiagonal matrix:

$$\begin{bmatrix} b_2 & c_2 & 0 & 0 & \cdots & 0 & 0 & 0 & 0 \\ a_3 & b_3 & c_3 & 0 & \cdots & 0 & 0 & 0 & 0 \\ 0 & a_4 & b_4 & c_4 & \cdots & 0 & 0 & 0 & 0 \\ \vdots & \vdots & \vdots & \vdots & \ddots & \vdots & \vdots & \vdots & \vdots \\ 0 & 0 & 0 & 0 & \cdots & a_{N-2} & b_{N-2} & c_{N-2} & 0 \\ 0 & 0 & 0 & 0 & \cdots & 0 & a_{N-1} & b_{N-1} & c_{N-1} \\ 0 & 0 & 0 & 0 & \cdots & 0 & 0 & a_N & b_N \end{bmatrix} \times \begin{bmatrix} q_{i(2)}^g \\ q_{i(3)}^g \\ q_{i(4)}^g \\ \vdots \\ q_{i(N-2)}^g \\ q_{i(N-1)}^g \\ q_{i(N)}^g \end{bmatrix} = \begin{bmatrix} w_2 \\ w_3 \\ \vdots \\ w_{N-2} \\ w_{N-1} \\ w_N \end{bmatrix}. \quad (A12)$$

Since the bottom mixing ratios are fixed as $q_{1(1)}^g = 3 \times 10^{-6}$ and $q_{2(1)}^g = 2.7 \times 10^{-5}$, we have

$$w_2 = -\frac{n_{atm(j)} q_{i(j,k)}^g}{\Delta t} + C_{i(j,k)} q_{i(j,k)}^{SVP} - P_{i(j)} + L_{i(j)}$$

$$- \frac{1}{4\Delta z^2} \left( -K_{zz(j+1)} n_{atm(j+1)} + K_{zz(j-1)} n_{atm(j-1)} + 4 K_{zz(j)} n_{atm(j)} \right)$$

$$\cdot q_{i(1)}^g.$$

(A13)

The top boundary conditions are zero fluxes:

$$a_N = \frac{1}{2\Delta z^2} \left( K_{zz(N)} n_{atm(N)} + K_{zz(N-1)} n_{atm(N-1)} \right), \quad (A14)$$

$$b_N = \left( -\frac{K_{zz(N)} n_{atm(N)} + K_{zz(N-1)} n_{atm(N-1)}}{2\Delta z^2} + C_{i(j,k)} - \frac{n_{atm(j)}}{\Delta t} \right), \quad (A15)$$

$$w_N = -\frac{n_{atm(N)} q_{i(Y,k)}^g}{\Delta t} + C_{i(N,k)} q_{i(N,k)}^{SVP} - P_{i(N)} + L_{i(N)} + \frac{0}{\Delta z}. \quad (A16)$$

We use Tridiagonal Matrix Algorithm (TDMA, Thomas, 1949) to solve this matrix and implicitly calculate the gas mixing ratio profiles.

2. Solving for condensates

The condensed species are solved explicitly by equation (A1). The condensed species are assumed to diffuse with the particles in the cloud layer:

$$S_{diff} = -\frac{\partial}{\partial z} \left( -K_{zz} n_{atm} \frac{\partial \frac{M_p}{n_{atm}}}{\partial z} \right) \left( \frac{M_i q_i^c}{\sum_i (M_i q_i^c)} \right) \frac{N_a}{M_i}, \quad (A17)$$

where $q_i^c$ profiles are from the last time step, $M_p$ is the cloud mass loading:

$$M_p = (M_1 q_1^c + M_2 q_2^c) \frac{n_{atm}}{N_a}, \quad (A18)$$

where the top boundaries are zero fluxes and $q_{i(1)}^c = 0$. We did not include the aqueous chemistry. The condensation rates of condensed species are equals to the opposite numbers of the gas condensation rates, respectively, for mass conservation:

$$S_{cond} = -C_i \left(q_i^g - q_i^{svp}\right). \tag{A19}$$

The sedimentation is obtained by

$$S_{sed} = \frac{\partial}{\partial z}(v q_i^c n_{atm}), \tag{A20}$$

where the top boundaries are zero fluxes. Because the condensed species is solved using an explicit rather than implicit method, the discretization and time marching are straightforward and not listed here.

At every timestep, the mixing ratios of the gases are solved implicitly. The condensates are solved explicitly in a sub-time step. After the explicit integration of the condensates, we feedback the condensation rate to the implicit gas solver for the next timestep. The model reaches the steady state after integrating the system for 800 Earth days. Running the nominal case on a single core (MacBook Pro, 2.3 GHz Intel Core i5), our Python code can reach an "approximate" steady state in about 15 mins using a large timestep ($4\times10^3$ s). After that, most quantities do not change by much. But if one wants to reach a very detailed, high-precision balance of fluxes at the cloud base, it requires a finer timestep and might take half an hour to reach the "real" steady state depending on the error tolerance level in the definition of the flux balance.

**Data Availability Statement**

Simulation data from our work are available online at https://doi.org/10.6084/m9.figshare.19129844.v1 (Dai, 2022).

**Acknowledgments**

This work is supported by International Program for Ph.D. Candidates at Sun Yat-Sen University to L.D., the China Scholarship Council Fellowship to W. D. S., and National Science Foundation (NSF) Grant AST1740921 to X. Z. We thank the two anonymous reviewers for their thorough and constructive comments to greatly improve this work.

Kerzhanovich, V. V., & Marov, M. Ya. (1983). The atmospheric dynamics of Venus according to Doppler measurements by the Venera entry probes. In *Venus* (pp. 766–775). Tucson, AZ: University of Arizona Press.

Knollenberg, R. G., & Hunten, D. M. (1980). The microphysics of the clouds of Venus: Results of the Pioneer Venus Particle Size Spectrometer Experiment. *Journal of Geophysical Research: Space Physics*, *85*(A13), 8039–8058. https://doi.org/10.1029/JA085iA13p08039

Kolodner, M. A., & Steffes, P. G. (1998). The Microwave Absorption and Abundance of Sulfuric Acid Vapor in the Venus Atmosphere Based on New Laboratory Measurements. *Icarus*, *132*(1), 151–169. https://doi.org/10.1006/icar.1997.5887

Kopparla, P., Seshadri, A., Imamura, T., & Lee, Y. J. (2020). A Recharge Oscillator Model for Interannual Variability in Venus' Clouds. *Journal of Geophysical Research: Planets*, *125*(11), e2020JE006568. https://doi.org/10.1029/2020JE006568

Krasnopolsky, V. A. (1985). Chemical composition of Venus clouds. *Planetary and Space Science*, *33*(1), 109–117. https://doi.org/10.1016/0032-0633(85)90147-3

Krasnopolsky, V. A. (2007). Chemical kinetic model for the lower atmosphere of Venus. *Icarus*, *191*(1), 25–37. https://doi.org/10.1016/j.icarus.2007.04.028

Krasnopolsky, V. A. (2012). A photochemical model for the Venus atmosphere at 47–112km. *Icarus*, *218*(1), 230–246. https://doi.org/10.1016/j.icarus.2011.11.012

Krasnopolsky, V. A. (2015). Vertical profiles of $H_2O$, $H_2SO_4$, and sulfuric acid concentration at 45–75km on Venus. *Icarus*, *252*, 327–333. https://doi.org/10.1016/j.icarus.2015.01.024

Krasnopolsky, V. A., & Parshev, V. A. (1981). Chemical composition of the atmosphere of Venus. *Nature*, *292*(5824), 610–613. https://doi.org/10.1038/292610a0

Krasnopolsky, V. A., & Pollack, J. B. (1994). $H_2O$-$H_2SO_4$ System in Venus' Clouds and OCS, CO, and $H_2SO_4$ Profiles in Venus' Troposphere. *Icarus*, *109*(1), 58–78. https://doi.org/10.1006/icar.1994.1077

Kulmala, M., & Laaksonen, A. (1990). Binary nucleation of water–sulfuric acid system: Comparison of classical theories with different $H_2SO_4$ saturation vapor pressures. *The Journal of Chemical Physics*, *93*(1), 696–701. https://doi.org/10.1063/1.459519

Lane, W. A., & Opstbaum, R. (1983). High altitude Venus haze from Pioneer Venus limb scans. *Icarus*, *54*(1), 48–58. https://doi.org/10.1016/0019-1035(83)90071-4

Lee, Y. J., Titov, D. V., Tellmann, S., Piccialli, A., Ignatiev, N., Pätzold, M., Häusler, B., Piccioni, G., & Drossart, P. (2012). Vertical structure of the Venus cloud top from the VeRa and VIRTIS observations onboard Venus Express. *Icarus*, *217*(2), 599–609. https://doi.org/10.1016/j.icarus.2011.07.001
48